\documentclass[10pt]{iopart}

\expandafter\let\csname equation*\endcsname\relax
\expandafter\let\csname endequation*\endcsname\relax
\usepackage{mathtools}

\usepackage{lipsum}
\usepackage{amsfonts}
\usepackage{graphicx}
\usepackage{epstopdf}
\usepackage{algorithmic}
\usepackage{cleveref}
\usepackage{siunitx}
\usepackage[caption=false]{subfig} 

\begin{document}
\title[Modeling Deformed Transmission Lines for Continuous Strain Sensing Applications]{Modeling Deformed Transmission Lines for Continuous Strain Sensing Applications}

\author{Stefan H.~Strub}

\address{Institute for Theoretical Physics, ETH Zurich, 8093, Zurich, Switzerland}
\address{LEONI Studer AG, Hohlstrasse 190, 8004, Zurich,
Switzerland}
\author{Lucas B\"ottcher}
\address{Institute for Theoretical Physics, ETH Zurich, 8093, Zurich, Switzerland}
\address{Center of Economic Research, ETH Zurich, 8092, Zurich, Switzerland}
\address{LEONI Studer AG, Hohlstrasse 190, 8004, Zurich,
Switzerland}
\ead{lucasb@ethz.ch}

\begin{abstract}
Transmission lines are essential components in various signal and power distribution systems. In addition to their main use as connecting elements, transmission lines can also be employed as continuous sensors for the measurement and detection of external influences such as mechanical strains and deformations. The measuring principle is based on deformation-induced changes of the characteristic impedance. Reflections of an injected test signal at resulting impedance mismatches can be used to infer applied deformations. To determine the effect of deformations on the characteristic impedance, we develop a numerical framework that allows us to solve Maxwell's equations for any desired transmission-line geometry over a wide frequency range. The proposed framework utilizes a staggered finite-difference Yee method on non-uniform grids to efficiently solve a set of decoupled partial differential equations that we derive from the frequency domain Maxwell equations. To test our framework, we compare simulation results with analytical predictions and corresponding experimental data. Our results suggest that the proposed numerical framework is able to capture experimentally observed deformation effects and may therefore be used in transmission-line-based deformation and strain sensing applications. Furthermore, our framework can also be utilized to simulate and study the electromagnetic properties of complex arrangements of conductor, insulator, and shielding materials.
\end{abstract}


\submitto{\MST}

\maketitle
\ioptwocol

\section{Introduction}
The transmission of electromagnetic waves is at the heart of many signal and power distribution applications. For wavelengths that are small compared to the conductor length, transmission lines such as coaxial cables or twisted pairs are used to limit radiative losses and reflections. In addition to their application in signal and power transmission systems, it is also possible to use transmission lines as continuous sensors for the measurement and detection of different environmental influences~\cite{LiThAb,zhu2019truly}. One important application is the measurement of mechanical strains and deformations to monitor the condition of civil engineering structures such as houses, bridges, and damns~\cite{LiThAb}. In contrast to discrete sensors (\textit{e.g.}, strain and displacement gauges), the advantage of continuous transmission-line sensors is the possibility to perform measurements at any location along a sensing line. The measuring principle in electrical transmission lines is based on two mechanisms: (i) Localized mechanical strains lead to transmission-line deformations and corresponding changes of the characteristic impedance. (ii) An injected test signal gets reflected at deformation-induced impedance mismatches and the information contained in signal reflections can be used to infer applied deformations.

Electrical transmission lines are, however, not the only possibility to realize continuous strain and deformation sensors in practice. Another approach is based on optical fibres and Bragg grating and Brillouin scattering methods. The idea behind the Bragg grating method is to use optical gratings within optical fibres and measure wave-length changes that correspond to certain deformations. The disadvantage of this method is that it is only quasi-continuous and requires a large number of gratings to achieve good resolution~\cite{sirkis1998using}. In the Brillouin scattering technique, frequency shifts of the scattered light relative to the incident light are used to infer deformations. Despite the fact that this method does not require discrete gratings, the possible resolution is limited to about $10~\text{cm}$~\cite{brown1999brillouin,murayama2004distributed}. In addition to the low spatial resolution, another drawback of optical methods is that they may not be suitable for cost-sensitive applications. 
Electrical transmission lines, on the other hand, can be fabricated more cost-efficiently and allow for resolutions of greater than $1~\text{cm}$~\cite{LiThAb}.

To measure deformations with electrical transmission lines, it is necessary to determine the effect of deformations on the characteristic impedance.
However, existing models only capture concentric transmission-line deformations and thus neglect any further shape-specific information~\cite{LiThAb}.
To accurately describe the effect of arbitrary deformations on the characteristic impedance, we develop a numerical framework to solve Maxwell's equations in the frequency domain for any desired transmission-line geometry over a wide frequency range. Our framework is based on a staggered finite-difference Yee method on non-uniform grids and solves a set of decoupled partial differential equations (PDEs) that we derive from the frequency domain Maxwell equations.
We test the ability of the proposed framework to characterize deformation effects in transmission lines by comparing simulation results with analytical predictions and corresponding experimental data.
\section{Transmission-line theory}
\label{sec:Telegraphers}
\begin{figure*}
  \centering
  \includegraphics[width=0.5\textwidth]{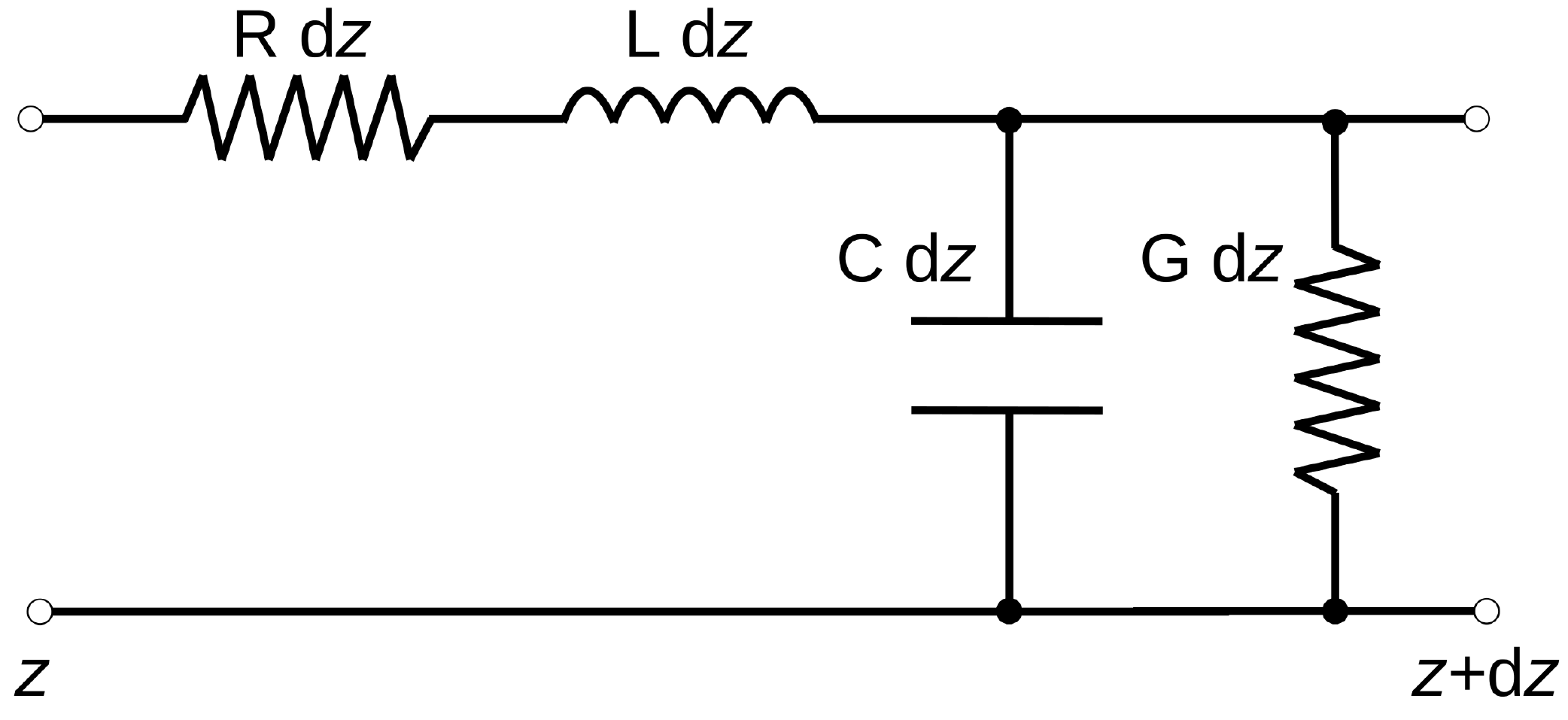}
  \caption{\textbf{Transmission line model.} The transmission line model is based on equivalent circuit representations of individual line segments where $R$, $L$, $C$, and $G$ are resistance, inductance, capacitance, and conductance per unit length, respectively. The direction of propagation of the incident electromagnetic wave is $z$.}
  \label{fig:transmissionline}
\end{figure*}
Before focusing on the development of our numerical framework for the study of deformed transmission lines, we summarize some concepts from transmission-line theory. In most transmission-line applications, the transversal electromagnetic (TEM) mode can be regarded as the dominant signal propagation mode~\cite{paul2008analysis}. In this mode, electric and magnetic fields lie in the transverse $(x,y)$ plane orthogonal to the direction of propagation $z$ of the incident electromagnetic wave. Based on the TEM assumption, it is possible to derive the telegrapher's equations from Maxwell's theory and describe the TEM wave propagation along a transmission line~\cite{chicone2016invitation,paul2008analysis}. In the frequency domain, these equations describe the transmission-line voltage $V(z)$ and current $I(z)$ at position $z$ and are given by
\begin{subequations}
\begin{align}
\frac{\mathrm{d} V\left(z\right)}{\mathrm{d} z}=-\left(R+ i \omega L\right) I\left(z\right)\,, \label{eq:telegraphers1} \\
\frac{\mathrm{d} I\left(z\right)}{\mathrm{d} z}=-\left(G+ i \omega C\right) V\left(z\right) \label{eq:telegraphers2}\,,
\end{align}
\end{subequations}
where $\omega=2 \pi f$ and $f$ is the corresponding frequency. The transmission-line parameters are $R=R(z,\omega)$, $L=L(z,\omega)$, $C=C(z,\omega)$, and $G=G(z,\omega)$ and denote resistance, inductance, capacitance, and conductance per unit length, respectively. Some important factors that influence these parameters are transmission-line geometry, conductivity and dielectric properties of the used materials, and frequency of the incident wave.

The telegrapher's equations may also be interpreted in terms of a sequence of equivalent circuits as we show in figure~\ref{fig:transmissionline}. Each segment is specified by the corresponding local transmission-line parameters and when put together they describe the whole transmission line. Instead of solving Maxwell's equations for a given transmission line and incident wave in three dimensions, the telegrapher's equations provide a way to describe the same problem by solving~\cref{eq:telegraphers1,eq:telegraphers2}. This approach permits a substantial reduction of computational effort. However, it also relies on the assumption of a TEM mode wave propagation that is only valid for uniform transmission lines consisting of perfect conductors surrounded by a homogeneous medium~\cite{paul2008analysis}. Still, even in the cases of lossy conductors, inhomogeneous surrounding media, and nonuniform cross-sections, the telegrapher's equations are used under the assumption that such effects only lead to negligible deviations from the TEM mode description (quasi-TEM assumption)~\cite{paul2008analysis}. We therefore simulate deformed transmission lines according to~\cref{eq:telegraphers1,eq:telegraphers2}.

In a transmission line, electromagnetic waves propagate with a phase velocity of $v=1/\sqrt{L C}$. If the surrounding medium is homogeneous with permeability $\mu$ and permittivity $\epsilon=\epsilon^\prime-i\,\epsilon^{\prime \prime}=\epsilon_0 \left(\epsilon_{\text{r}}^\prime-i\,\epsilon_{\text{r}}^{\prime\prime}\right)$, the phase velocity is $v=1/\sqrt{\mu \epsilon^\prime}$~\cite{paul2008analysis}. The imaginary part of the permittivity accounts for dielectric losses and $\epsilon_0$ is the vacuum permittivity. In most transmission line applications, the permeability $\mu=\mu_0 \mu_{\text{r}}$ is equal to the vacuum permeability $\mu_0$, because common conductors (\textit{e.g.}, copper and aluminum) exhibit a relative permeability of $\mu_{\text{r}}\approx 1$.  

For each transmission-line segment, the characteristic impedance is~\cite{paul2008analysis}
\begin{equation}
Z_0\left(\omega\right) = \sqrt{\frac{R+ i \omega L}{G+i \omega C}}
\label{eq:impedance}
\end{equation}
and therefore a function of the transmission-line parameters.
At an interface between a line segment with impedance $Z_1$ and another one with impedance $Z_2$, the corresponding reflection coefficient~\cite{paul2008analysis}
\begin{equation}
\Gamma_{12} = \frac{Z_2-Z_1}{Z_2+Z_1}
\label{eq:reflection_coeff}
\end{equation}
describes the ratio of the reflected and the incident wave. No reflections occur for transmission lines with equal impedances along the line (matched transmission lines). To describe the influence of transmission-line deformations on characteristic impedance and signal propagation, we have to determine the corresponding transmission-line parameters. Only in certain cases such as for coaxial geometries, is it possible to find closed analytical solutions. For general deformations, it is necessary to utilize numerical methods. 

Under the quasi-TEM assumption, we first determine the electric and magnetic fields $\mathbf{E}\left( z,\omega\right)$ and $\mathbf{B}\left(z,\omega\right)$ for a transmission-line segment at position $z$. Based on the obtained electromagnetic fields, we then compute the transmission-line parameters for the considered segment. Specifically, let $\Omega  \subseteq  \mathbb{R}^2$ be the cross section of a certain transmission-line segment at position $z$. The cross section $\Omega$ shall include the surrounding medium up to a certain distance where the magnetic field is sufficiently small. We denote the current density in $z$ direction by $J_z\left(z,\omega\right)$ and describe resistive losses by~\cite{jackson1999classical} 
\begin{equation}
P\left(z,\omega\right) = \int_\Omega \frac{{J_z\left(z,\omega\right)}^2}{\sigma} \mathrm{d} A\,,
\label{eq:jouleheating}
\end{equation}
where $\sigma$ denotes the electric conductivity. Based on \cref{eq:jouleheating}, the resistance per unit length is
\begin{equation}
R\left(z,\omega\right) = \frac{P\left(z,\omega\right)}{I^2}\,,
\label{eq:R}
\end{equation}
where $I$ is the current that flows through the transmission line. Next, we compute the inductance with the help of the total magnetic energy averaged over one cycle~\cite{jackson1999classical}
\begin{equation}
\bar{W}_m\left(z,\omega\right) = \frac{1}{4} \int_{\Omega} \mathbf{H}\left(z,\omega\right)\cdot \mathbf{B}^\ast\left(z,\omega\right)\, \mathrm{d} A\,,
\label{eq:magneticenergy}
\end{equation}
where $\mathbf{H}\left(z,\omega\right)=\mathbf{B}\left(z,\omega\right)/\mu $. The inductance per unit length is~\cite{jackson1999classical}
\begin{equation}
L\left(z,\omega\right) = \frac{4 \bar{W}_m\left(z,\omega\right)}{I^2}\,.
\label{eq:L}
\end{equation}
Similarly, we determine the capacitance by considering the total electric energy averaged over one cycle~\cite{jackson1999classical}
\begin{equation}
\bar{W}_e\left(z,\omega\right) = \frac{1}{4} \int_{\Omega} \mathbf{E}\left(z,\omega\right)\cdot \mathbf{D}^*\left(z,\omega\right) \mathrm{d} A\,,
\label{eq:electricenergy}
\end{equation}
where $\mathbf{D}\left(z,\omega\right)=\mathbf{E}\left(z,\omega\right)/\epsilon^\prime$. We obtain the capacitance per unit length through~\cite{jackson1999classical}
\begin{equation}
C\left(z,\omega\right) = \frac{4 \bar{W}_e\left(z,\omega\right)}{V^2}\,,
\label{eq:C}
\end{equation}
where $V$ is the voltage difference between inner and outer conductor. We note that electric field and capacitance are frequency-independent until currents that flow through the dielectric material become relevant. This is only the case for frequencies that are not captured by the quasi-TEM approximation. Another possibility to compute the capacitance is to consider the total charge $Q$. Therefore, let $\Omega' \subseteq \mathbb{R}^2$ be the area that covers all conductors of the same voltage level. The total charge is
\begin{align}
\begin{split}
Q\left(z,\omega\right) &= \epsilon^\prime \int_{\Omega'} \rho \, \mathrm{d}A \\
 &= \epsilon_0 \int_{\partial \Omega'} \epsilon_{\text{r}}^\prime \mathbf{E}\cdot \mathbf{n}' \, \mathrm{d} l\,,
\end{split}
\label{eq:total_charge}
\end{align}
where $\rho$ is the charge density. The relative permittivity $\epsilon_{\text{r}}^\prime$ may be position-dependent. The second step in \cref{eq:total_charge} follows from Gauss' law, where $\mathbf{n}'$ is perpendicular to the boundary $\partial \Omega'$ of $\Omega'$. The capacity per unit length is
\begin{equation}
C\left(z,\omega\right) = \frac{Q\left(z,\omega\right)}{V}\,.
\end{equation}
For a homogeneous dielectric medium, the conductance per unit length is~\cite{paul2008analysis}
\begin{equation}
G\left(z,\omega\right) = \omega \tan\left(\delta \right)   C\left(z,\omega\right)\,,
\label{eq:Gapprox}
\end{equation}
where 
\begin{equation}
\tan \left(\delta\right)=\frac{\omega \epsilon^{\prime \prime}- \sigma_{\text{dielectric}}}{\omega \epsilon^\prime}
\end{equation}
denotes the loss tangent. In practice, dielectric losses dominate and $\tan\left(\delta\right)=\epsilon^{\prime\prime}/\epsilon^\prime$. To fully describe the dielectric material, it is necessary to determine the frequency dependence of $\tan\left(\delta\right)$. For typical dielectric materials that are used in transmission line applications, the frequency dependence is almost constant over a certain frequency range~\cite{paul2008analysis}. In the more general case of a non-homogeneous dielectric medium, the conductance per unit length is~\cite{paul2008analysis}
\begin{equation}
G\left(z,\omega\right) = \frac{I_t\left(z,\omega\right)}{V}\,,
\end{equation}
where $I_t$ is the total transverse conduction current per unit line length between the conductors. Integrating the current density $\mathbf{J} = \left(\sigma - \omega \epsilon^{\prime \prime}\right) \mathbf{E}+\mathbf{J}^s$ over $\partial \Omega'$ yields the total transverse conduction current and the conductance per unit length
\begin{equation}
G\left(z,\omega\right) = \frac{1}{V} \int_{\partial \Omega'} \mathbf{J}\cdot \mathbf{n}' \, \mathrm{d}l\,,
\label{eq:G}
\end{equation}
where $\mathbf{J}^s$ in $\mathbf{J}$ is the externally applied source current.
\section{Potential formulation of Maxwell's equations}
\label{sec:Formulation}
After having outlined the basic strategy of how to describe transmission lines of arbitrary cross sections in section~\ref{sec:Telegraphers}, we now determine the electromagnetic fields of a given transmission-line segment~\cite{OldenburgFast,haber2001fast}. Based on~\cref{eq:R,eq:L,eq:C,eq:G}, it is then possible to obtain the corresponding transmission-line parameters.

We begin with a brief summary of the potential formulation of Maxwell's equations as proposed in~\cite{OldenburgFast,haber2001fast}. Maxwell's equations in the frequency domain are
\begin{align}
\nabla\times\mathbf{E} - i \omega \mu \mathbf{H}  &=  0\,, \label{eq:max1}\\
\nabla\times\mathbf{H} - \left(\sigma - i \omega \epsilon\right) \mathbf{E}  &=  \mathbf{J}^s\,, \label{eq:max2}\\
\nabla \cdot \left(\epsilon \mathbf{E}\right)  &=  \rho\,, \label{eq:max3}\\
\nabla \cdot \left(\mu \mathbf{H}\right)  &=  0\,. \label{eq:max4}
\end{align}
We divide \cref{eq:max1} by $\mu$, take the curl, and substitute \cref{eq:max2} into the resulting expression to obtain

\begin{equation}
\nabla\times \left(\mu^{-1} \nabla\times\mathbf{E} \right) - i \omega \hat{\sigma} \mathbf{E} = i \omega \mathbf{J}^s\,,
\label{eq:PDE E}
\end{equation}
where
\begin{equation}
\hat{\sigma} \coloneqq \sigma - i \omega \epsilon
\end{equation}
is the generalized conductivity.
Moreover, a decomposition of $\mathbf{E}$ into the vector potential $\mathbf{A}$ and the scalar potential $\Phi$ yields
\begin{equation}
\mathbf{E} = \mathbf{A} + \nabla \Phi \label{eq:decompose E}\,.
\end{equation}
The vector potential satisfies the Coulomb gauge condition
\begin{equation}
\nabla\cdot \mathbf{A} = 0\,.
\label{eq:gauge}
\end{equation}
Substituting \cref{eq:decompose E} into \cref{eq:PDE E} leads to
\begin{equation}
\nabla\times\nabla\times\mathbf{A} - i \omega \mu \hat{\sigma} \left(\mathbf{A} + \nabla \Phi \right)= i \omega \mu \mathbf{J}^s\,,
\label{eq:PDE_A_complex}
\end{equation}
where we assumed a constant permeability $\mu$. Using the Coulomb gauge condition of \cref{eq:gauge}, simplifies \cref{eq:PDE_A_complex} to
\begin{equation}
\nabla^2\mathbf{A} + i \omega \mu \hat{\sigma} \left(\mathbf{A} + \nabla \Phi \right)= - i \omega \mu \mathbf{J}^s
\label{eq:PDE A phi}
\end{equation}
and taking the divergence of \cref{eq:PDE A phi} yields
\begin{equation}
\nabla\cdot \left[\hat{\sigma} \left(\mathbf{A} + \nabla \Phi \right) \right] = - \nabla\cdot \mathbf{J}^s\,.
\label{eq:PDE all}
\end{equation}
As described in section~\ref{sec:Telegraphers}, under the quasi-TEM assumption we consider the transmission line to be composed of individual segments whose electromagnetic fields lie in the transverse $(x,y)$ plane. Therefore, the solution of \cref{eq:PDE A phi,eq:PDE all} can be obtained with a substantial reduction in computational effort. Under the quasi-TEM assumption, the resulting PDEs are
\begin{subequations}
\begin{align}
\left(\partial_x^2 + \partial_y^2\right) A_x + i \omega \mu \hat{\sigma} \left(A_x + \partial_x{\Phi} \right) &= - i \omega \mu J^{s}_x\,, \label{eq:A_x1}\\
\left(\partial_x^2 + \partial_y^2\right) A_y + i \omega \mu \hat{\sigma} \left(A_y + \partial_y{\Phi} \right)&= - i \omega \mu J^{s}_y\,, \label{eq:A_x2}\\
\left(\partial_x^2 + \partial_y^2\right) A_z + i \omega \mu \hat{\sigma} A_z &= - i \omega \mu J^{s}_z\,, \label{eq:A_z}\\
\partial_x{\left[\hat{\sigma} \left( A_x + \partial_x{\Phi} \right)\right]} + \partial_y{\left[\hat{\sigma} \left( A_y + \partial_y{\Phi}\right)\right]} &= -\partial_x J^{s}_x \nonumber \\
&- \partial_y J^{s}_y\,. \label{eq:A_x3}
\end{align}
\label{eq:PDE split}
\end{subequations}
We note that $\Phi$ dropped out of \cref{eq:A_z}. As a result $A_z$ is decoupled from $A_x, A_y$, and $\Phi$.
To determine capacitance and conductance for different frequencies, we have to solve \cref{eq:A_x1,eq:A_x2,eq:A_z,eq:A_x3} only once since the capacitance as defined in~\cref{eq:C} exhibits no frequency dependence and the conductance scales linearly with $\omega$ according to \cref{eq:G}~\cite{paul2008analysis}. In section~\ref{sec:electr_magn_fields}, we outline that it is sufficient to solve \cref{eq:A_z} to determine the frequency dependence of the magnetic field, and thus of the resistance and inductance of a certain transmission-line segment. This again leads to a substantial reduction of computational effort compared to the case in which we have to solve the whole set of PDEs given by~\cref{eq:A_x1,eq:A_x2,eq:A_z,eq:A_x3}.
According to \cite{OldenburgFast}, possible boundary conditions for the solution of \cref{eq:PDE A phi} are
\begin{subequations}
\begin{align}
\left(\nabla \times \mathbf{A} \right) \times \mathbf{n}|_{\partial \Omega} &= 0\,, \\
\mathbf{A} \cdot \mathbf{n}|_{\partial \Omega} &= 0\,, \\
\nabla \Phi |_{\partial \Omega}\cdot \mathbf{n}&= 0\,, \label{eq:Phi1}\\
\int_{\Omega} \Phi \mathrm{d}A &= 0 \label{eq:Phi2}\,,
\end{align}
\label{eq:BC}
\end{subequations}
where $\mathbf{n}$ is the unit normal vector on the boundary $\partial \Omega$. In the case of transmission lines, the inner and outer conductors can be assumed to exhibit potentials of the same absolute value relative to the respective area, but with opposite signs. Therefore, the boundary conditions given by~\cref{eq:Phi1,eq:Phi2} are already satisfied.
\section{Discretization}
\label{sec:Discretization}
\begin{figure}
  \centering
  \includegraphics[width=0.37\textwidth]{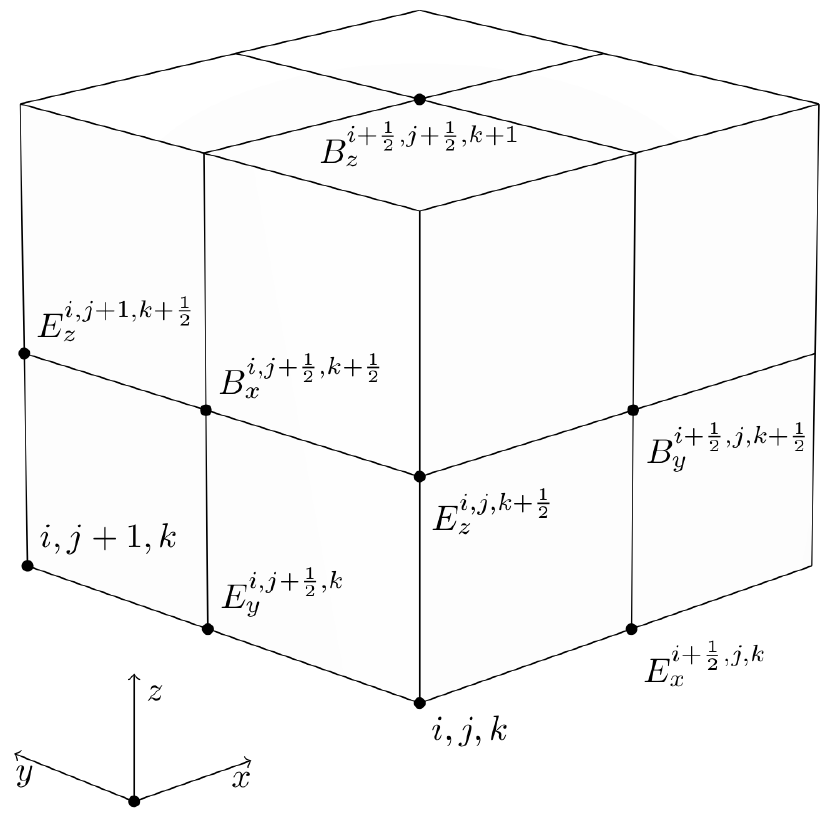}
  \caption{\textbf{Electromagnetic field discretization.}  Position of the discretized electromagnetic field components according to ~\cite{YeeNum}.}
  \label{fig:yee_grid}
\end{figure}
To solve~\cref{eq:A_x1,eq:A_x2,eq:A_z,eq:A_x3} for arbitrary transmission-line cross sections, we employ a staggered finite-difference scheme on non-uniform grids~\cite{YeeNum,haber2001fast,OldenburgFast}. Specifically, we use a method that is based on central differences at midpoints. That is, we compute derivatives between two adjacent grid points using a forward-difference scheme and consider the resulting derivative values to be located at the corresponding midpoints. In this way, boundaries between different conductors and dielectric materials can be resolved well. 

\begin{figure}
  \centering
  \subfloat[Electromagnetic field components in the plane.]{\label{fig:yee_grid_square}\includegraphics[width=0.5\columnwidth,trim=0.3cm 1.2cm 1.5cm 1.5cm]{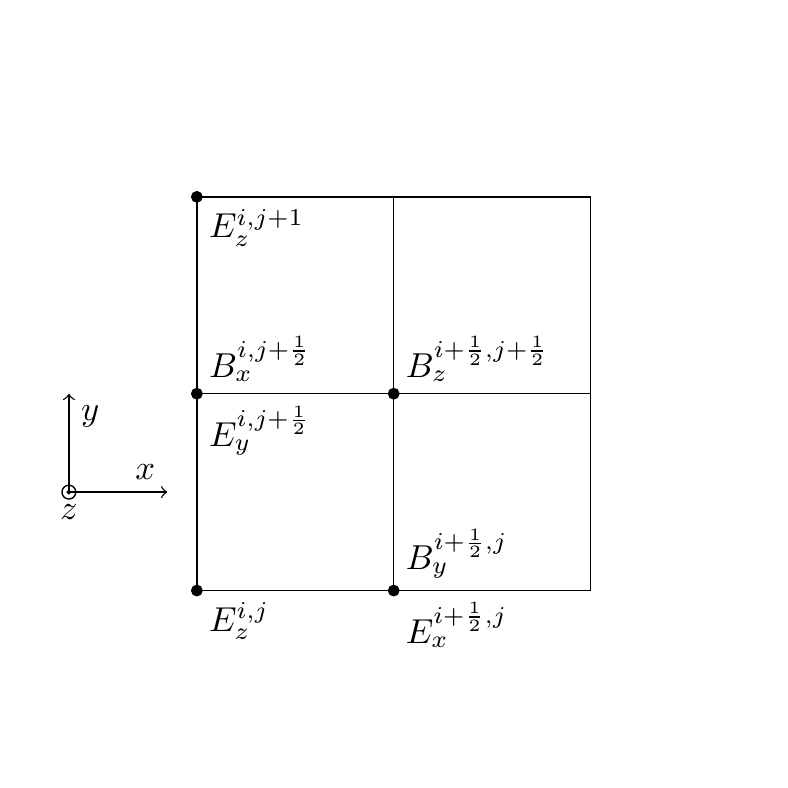}}
  \subfloat[Potential components in the plane.]{\label{fig:yee_grid_potentials}\includegraphics[width=0.5\columnwidth,trim=0.3cm 1.2cm 1.5cm 1.5cm]{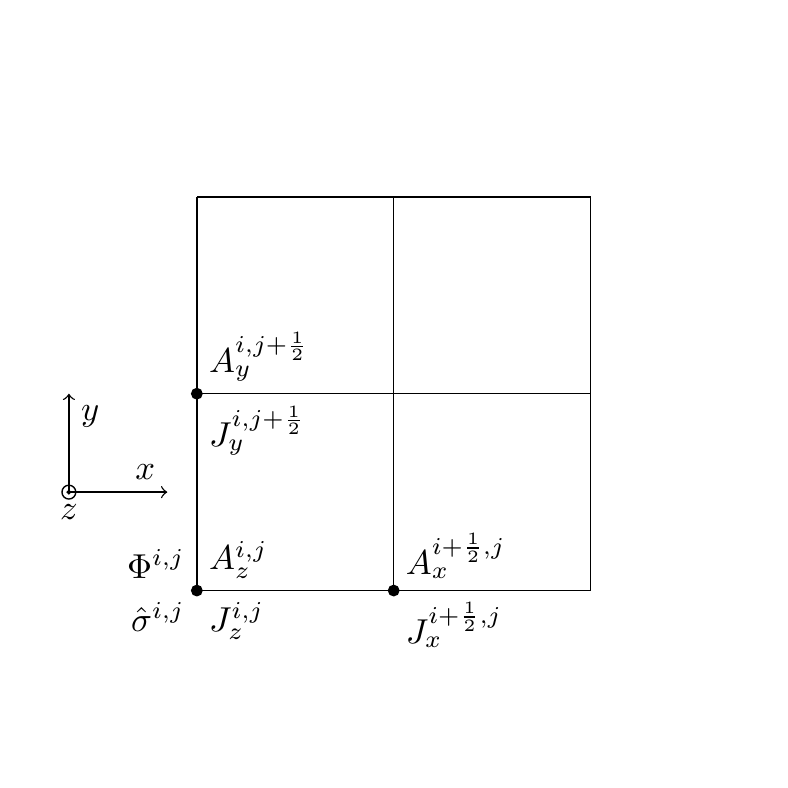}}
  \caption{\textbf{Reduction to a two-dimensional problem.} Under the quasi-TEM assumption, it is possible to consider a two-dimensional version of the Yee grid.}
  \label{fig:yee_grid_planes}
\end{figure}

We discretize the electromagnetic fields according to the Yee method (see figure~\ref{fig:yee_grid})~\cite{YeeNum}. Here and in the subsequent sections, we use $F_x^{i,j}$ as a shorthand notation for the the $x$-component $F_x \left(x_i, y_j\right)$ of a vector field $\mathbf{F}$ at position $\left(x_i, y_j\right)$, and similarly for the $y$ and $z$ components of $\mathbf{F}$. Instead of solving the three-dimensional problem (see figure~\ref{fig:yee_grid}), we consider a two-dimensional representation of the Yee grid (see figure~\ref{fig:yee_grid_planes}).
This reduction to two dimensions is a consequence of the quasi-TEM assumption that we described in section~\ref{sec:Formulation}. We show the discretization of the electromagnetic potentials and corresponding parameters in figure~\ref{fig:yee_grid_potentials}.
\subsection{Non-uniform grid}
\begin{figure*}
  \centering
  \includegraphics{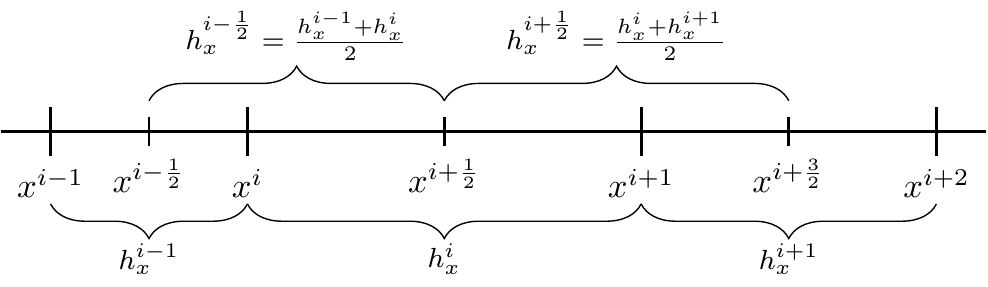}
  \caption{\textbf{Non-uniform grid.} An illustration of a non-uniform discretization along the $x$-axis.}
  \label{fig:tikz_x_line}
\end{figure*}

One possibility to numerically solve~\cref{eq:A_x1,eq:A_x2,eq:A_z,eq:A_x3}, is to consider equal grid spacings in a two-dimensional Yee grid (see figure \ref{fig:yee_grid_planes}). For perfectly coaxial geometries, such an approach is suitable to resolve the electromagnetic fields since they vanish outside the transmission line. However, for arbitrary geometries, it is important to resolve far field contributions of the magnetic field to correctly compute the inductance according to \cref{eq:magneticenergy,eq:L}. We therefore employ a non-uniform Yee discretization of the computational domain $\Omega\subseteq \mathbb{R}^2$ (see figure~\ref{fig:tikz_x_line}) and denote the difference between grid points $x^{i}$ ($y^{i}$) and $x^{i+1}$ ($y^{i+1}$) by $h_x^{i}$ ($h_y^{i}$).

We first focus on the non-uniform Yee discretization of all derivatives occurring in~\cref{eq:A_x1,eq:A_x2,eq:A_z,eq:A_x3} and then present the fully discretized version of the considered PDEs. We only describe the discretization along the $x$-axis bearing in mind that the same steps also apply in $y$-direction. For the discretization of $\partial_x \Phi$, we consider central differences at midpoints and thus compute the derivative between $x^i$ and $x^{i+1}$ at position $x^{i+\frac{1}{2}}$ (see figure~\ref{fig:tikz_x_line}). We obtain
\begin{equation}
\left( \partial_x \Phi \right)^{i+\frac{1}{2},j} = \frac{\Phi^{i+1,j} - \Phi^{i,j}}{h_x^i}\,.
\end{equation}
In the next step, we discretize $\hat{\sigma} A_x$ and consider $\hat{\sigma}^{i,j}$ to be located at the same position as $\Phi^{i,j}$ and $J_z^{i,j}$ (see figure~\ref{fig:yee_grid_potentials}). We have to determine $\hat{\sigma}^{i+\frac{1}{2},j}$ since the vector field component $A_x^{i+\frac{1}{2}}$ is located at $x^{i+\frac{1}{2}}$. To do so, we compute the harmonic mean
\begin{equation}
\hat{\sigma}^{i+\frac{1}{2},j} = 2 \left( \frac{1}{\hat{\sigma}^{i,j}} + \frac{1}{\hat{\sigma}^{i+1,j}} \right)^{-1}
\end{equation}
and obtain
\begin{equation}
\hat{\sigma}^{1+\frac{1}{2},j}  A_x^{i+\frac{1}{2},j} = 2 \left( \frac{1}{\hat{\sigma}^{i,j}} + \frac{1}{\hat{\sigma}^{i+1,j}} \right)^{-1}  A_x^{i+\frac{1}{2},j}
\end{equation}
and 
\begin{equation}
\left[\partial_x \left( \hat{\sigma}  A_x \right) \right]^{i,j} = \frac{\hat{\sigma}^{i+\frac{1}{2},j} A_x^{i+\frac{1}{2},j}- \hat{\sigma}^{i-\frac{1}{2},j} A_x^{i-\frac{1}{2},j}}{h_x^{i-\frac{1}{2}}}\,.
\end{equation}
For computing the second derivatives of the electromagnetic potentials, we consider the Taylor expansions
\begin{equation}
 f^{i-1,j} = f^{i,j} - h_x^{i-1} \partial_x f^{i,j} + \frac{1}{2!}\left(h_x^{i-1}\right)^2 \partial^2_x f^{i,j} - \mathcal{O}\left(\left(h_x^{i-1}\right)^3\right)
\end{equation}
and
\begin{equation}
 f^{i+1,j} = f^{i,j} + h_x^{i} \partial_x f^{i,j} + \frac{1}{2!}\left(h_x^{i}\right)^2 \partial^2_x f^{i,j}+ \mathcal{O}\left(\left(h_x^{i}\right)^3\right)
\end{equation}
of a function $f$. We eliminate $\partial_x f^{i,j}$ to obtain
\begin{align}
\begin{split}
\left(\partial^2_x f \right)^{i,j} &= \frac{2 f^{i-1,j}}{h_x^{i-1} \left(h_x^{i-1}+h_x^{i}\right)} - \frac{2 f^{i,j}}{h_x^{i-1} h_x^{i}} \\
&+ \frac{2 f^{i+1,j}}{h_x^{i} \left(h_x^{i-1}+h_x^{i}\right)} +  \mathcal{O}\left(h_x^{i-1}-h_x^i\right)
\end{split}
\end{align}
and similarly for a vector potential component
\begin{equation}
\begin{split}
\left(\partial^2_x A_y \right)^{i,j+\frac{1}{2}} &= \frac{2 A_y^{i-1,j+\frac{1}{2}}}{h_x^{i-1} \left(h_x^{i-1}+h_x^{i}\right)} - \frac{2 A_y^{i,j+\frac{1}{2}}}{h_x^{i-1} h_x^{i}} \\
&+ \frac{2 A_y^{i+1,j+\frac{1}{2}}}{h_x^{i} \left(h_x^{i-1}+h_x^{i}\right)} +  \mathcal{O}\left(h_x^{i-1}-h_x^i\right)\,.
\end{split}
\label{eq:secondDerivative}
\end{equation}
Since the vector potentials are located between grid points, we replace $h_x^{i,j}$ by $h_x^{i+\frac{1}{2},j} = \left(h_x^{i,j}+h_x^{i+1,j}\right)/2$ and, for the second derivative of the vector potential, obtain
\begin{align}
\begin{split}
\left(\partial^2_x A_x \right)^{i+\frac{1}{2},j} &= \frac{2 A_x^{i-\frac{1}{2},j}}{h_x^{i-\frac{1}{2}} \left(h_x^{i-\frac{1}{2}}+h_x^{i+\frac{1}{2}}\right)} - \frac{2 A_x^{i+\frac{1}{2},j}}{h_x^{i-\frac{1}{2}} h_x^{i+\frac{1}{2}}} \\
&+ \frac{2 A_x^{i+\frac{3}{2},j}}{h_x^{i+\frac{1}{2}} \left(h_x^{i-\frac{1}{2}}+h_x^{i+\frac{1}{2}}\right)}\,.
\end{split}
\end{align}
Furthermore, we compute the term $\partial_x{\left(\hat{\sigma} \partial_x{\Phi} \right)}$ in \cref{eq:A_x3} by considering central differences at midpoints:
\begin{align}
\begin{split}
\partial_x{\left(\hat{\sigma} \partial_x{\Phi} \right)} &= \frac{\hat{\sigma}^{i-\frac{1}{2},j} \left(\Phi^{i-1,j} - \Phi^{i,j}\right)}{h_x^{i-\frac{1}{2}} h_x^{i-1}} \\
&- \frac{\hat{\sigma}^{i+\frac{1}{2},j} \left(\Phi^{i,j}- \Phi^{i+1,j}\right)}{h_x^{i-\frac{1}{2}}h_x^{i}}\,.
\end{split}
\end{align}
Finally, we can formulate~\cref{eq:A_x1,eq:A_x2,eq:A_z,eq:A_x3} in terms of a staggered finite-difference Yee discretization on a non-uniform grid:
\begin{align}
\begin{split}
&\frac{2 A_x^{i-\frac{1}{2},j}}{h_x^{i-\frac{1}{2}} \left(h_x^{i-\frac{1}{2}}+h_x^{i+\frac{1}{2}}\right)} - \frac{2 A_x^{i+\frac{1}{2},j}}{h_x^{i-\frac{1}{2}} h_x^{i+\frac{1}{2}}} \\
&+ \frac{2 A_x^{i+\frac{3}{2},j}}{h_x^{i+\frac{1}{2}} \left(h_x^{i-\frac{1}{2}}+h_x^{i+\frac{1}{2}}\right)} + \frac{2 A_x^{i+\frac{1}{2},j-1}}{h_y^{j-1} \left(h_y^{j-1}+h_y^{j}\right)} \\
&- \frac{2 A_x^{i+\frac{1}{2},j}}{h_y^{j-1} h_y^{j}} + \frac{2 A_x^{i+\frac{1}{2},j+1}}{h_y^{j} \left(h_y^{j-1}+h_y^{j}\right)} \\
& + i \omega \mu \hat{\sigma}^{i+\frac{1}{2},j} \left( A_{x}^{i+\frac{1}{2},j} + \frac{\Phi^{i,j} + \Phi^{i+1,j}}{h_x^i}\right)   = -i \omega \mu J_{s,x}^{i+\frac{1}{2},j},
\label{eq:discr_1}
\end{split}
\end{align}
\begin{align}
\begin{split}
&\frac{2 A_y^{i-1,j+\frac{1}{2}}}{h_x^{i-1} \left(h_x^{i-1}+h_x^{i}\right)} - \frac{2 A_y^{i,j+\frac{1}{2}}}{h_x^{i-1} h_x^{i}} \\
&+ \frac{2 A_y^{i+1,j+\frac{1}{2}}}{h_x^{i} \left(h_x^{i-1}+h_x^{i}\right)}+ \frac{2 A_y^{i,j-\frac{1}{2}}}{h_y^{j-\frac{1}{2}} \left(h_y^{j-\frac{1}{2}}+h_y^{j+\frac{1}{2}}\right)} \\
&- \frac{2 A_y^{i,j+\frac{1}{2}}}{h_y^{j-\frac{1}{2}} h_y^{j+\frac{1}{2}}} + \frac{2 A_y^{i,j+\frac{3}{2}}}{h_y^{j+\frac{1}{2}} \left(h_y^{j-\frac{1}{2}}+h_y^{j+\frac{1}{2}}\right)} \\ 
&+ i \omega \mu \hat{\sigma}^{i,j+\frac{1}{2}} \left( A_{y}^{i,j+\frac{1}{2}} + \frac{\Phi^{i,j} + \Phi^{i,j+1}}{h_y^j}\right)   = -i \omega \mu J_{s,y}^{i,j+\frac{1}{2}},
\label{eq:discr_2}
\end{split}
\end{align}
\begin{align}
\begin{split}
&\frac{2 A_z^{i-1,j}}{h_x^{i-1} \left(h_x^{i-1}+h_x^{i}\right)} - \frac{2 A_z^{i,j}}{h_x^{i-1} h_x^{i}} + \frac{2 A_z^{i+1,j}}{h_x^{i} \left(h_x^{i-1}+h_x^{i}\right)} \\ 
&+ \frac{2 A_z^{i,j-1}}{h_y^{j-1} \left(h_y^{j-1}+h_y^{j}\right)} - \frac{2 A_z^{i,j}}{h_y^{j-1} h_y^{j}} + \frac{2 A_z^{i,j+1}}{h_y^{j} \left(h_y^{j-1}+h_y^{j}\right)} \\ 
&+ i \omega \mu \hat{\sigma}^{i,j} A_{z}^{i,j}   = -i \omega \mu J_{s,z}^{i,j}\,,
\label{eq:discr_3}
\end{split}
\end{align}
and
\begin{align}
\begin{split}
&\frac{\hat{\sigma}^{i+\frac{1}{2},j} A_x^{i+\frac{1}{2},j}- \hat{\sigma}^{i-\frac{1}{2},j} A_x^{i-\frac{1}{2},j}}{h_x^{i-\frac{1}{2}}} + \frac{\hat{\sigma}^{i,j+\frac{1}{2}} A_y^{i,j+\frac{1}{2}}- \hat{\sigma}^{i,j-\frac{1}{2}} A_y^{i,j-\frac{1}{2}}}{h_y^{j-\frac{1}{2}}} \\
&+\frac{\hat{\sigma}^{i+\frac{1}{2},j} \left(\Phi^{i+1,j} - \Phi^{i,j}\right)}{h_x^{i-\frac{1}{2}}h_x^{i}}  - 
\frac{\hat{\sigma}^{i-\frac{1}{2},j} \left(\Phi^{i,j} - \Phi^{i-1,j}\right)}{h_x^{i-\frac{1}{2}} h_x^{i-1}} \\
&+ \frac{\hat{\sigma}^{i,j+\frac{1}{2}} \left(\Phi^{i,j+1} - \Phi^{i,j}\right)}{h_y^{j-\frac{1}{2}}h_y^{j}}  - 
\frac{\hat{\sigma}^{i,j-\frac{1}{2}} \left(\Phi^{i,j} - \Phi^{i,j-1}\right)}{h_y^{j-\frac{1}{2}} h_y^{j-1}} \\
& =  \frac{J_{s,x}^{i-\frac{1}{2},j} - J_{s,x}^{i+\frac{1}{2},j}}{h_x^{i-\frac{1}{2}}} + \frac{J_{s,y}^{i,j-\frac{1}{2}} - J_{s,y}^{i,j+\frac{1}{2}}}{h_y^{j-\frac{1}{2}}}\,.
\label{eq:discr_4}
\end{split}
\end{align}
Our approach extends \cite{OldenburgFast} by considering a non-uniform discretization of the potential formulation of Maxwell's equations as defined by \cref{eq:A_x1,eq:A_x2,eq:A_z,eq:A_x3}.

To numerically solve \cref{eq:discr_1,eq:discr_2,eq:discr_3,eq:discr_4}, we rewrite the discretized PDEs in terms of a linear system of equations
\begin{equation}
D \mathbf{A}'=\mathbf{J}'\,,
\label{eq:system}
\end{equation}
where $\mathbf{J}'=- i \omega \mu\left(J_{s,x},  J_{s,y}, J_{s,z}, \nabla \mathbf{J_s}/(i \omega \mu)\right)$, $\mathbf{A}'=\left(A_x,A_y,A_z,\Phi\right)$, and

\begin{equation*}
D=
\begin{pmatrix}
\Delta + i \omega \mu \hat{\sigma} & 0 & 0 & i \omega \mu \hat{\sigma} \partial_x  \\
0 & \Delta + i \omega \mu \hat{\sigma}  & 0 & i \omega \mu \hat{\sigma} \partial_y \\
0 & 0  & \Delta + i \omega \mu \hat{\sigma} & 0\\\partial_x\hat{\sigma} & \partial_y\hat{\sigma} & 0 &  \nabla \hat{\sigma} \nabla
\end{pmatrix}\,.
\end{equation*}
We solve \cref{eq:system} with a sparse linear system solver and consider $J_{s,z}$ to be the only non-vanishing source current component.
\subsection{Computing electromagnetic fields}
\label{sec:electr_magn_fields}
After having determined the electromagnetic potentials by solving~\cref{eq:system}, we now compute the corresponding electromagnetic fields. The electric field $\mathbf{E}$ as defined by \cref{eq:decompose E} is
\begin{equation}
\mathbf{E}^{i,j} = \mathbf{A}^{i,j} + \begin{pmatrix}
\frac{\Phi^{i+1,j} - \Phi^{i,j}}{h_x^i} \\ \frac{\Phi^{i,j+1} - \Phi^{i,j}}{h_y^j} \\ 0
\end{pmatrix}\,.
\label{eq:E field}
\end{equation}
Based on \cref{eq:max1}, the magnetic field is
\begin{equation}
\mathbf{B} = \mu \mathbf{H} = \frac{1}{i \omega} \nabla\times\mathbf{E} = \frac{1}{i \omega} 
 \begin{pmatrix}
\partial_y E_z \\ - \partial_x E_z \\ \partial_x E_y - \partial_y E_x 
\end{pmatrix}
\end{equation}
and thus
\begin{equation}
\mathbf{B}^{i,j} = \frac{1}{i \omega} 
\begin{pmatrix}
\frac{E_z^{i,j+1} - E_z^{i,j}}{h_y^j} \\ - \frac{E_z^{i+1,j} - E_z^{i,j}}{h_x^i} \\ \frac{E_y^{i+1,j} - E_y^{i,j}}{h_x^i} - \frac{E_x^{i,j+1} - E_x^{i,j}}{h_y^j}
\end{pmatrix}\,.
\label{eq:B field}
\end{equation}
We only have to compute the electric field for one frequency, because the capacitance is frequency-independent and the conductance scales linearly with $\omega$ according to \cref{eq:G}. For all other frequencies, it is sufficient to numerically solve
\begin{equation}
\left( \Delta + i \omega \mu \hat{\sigma} \right) A_z  = - i \omega \mu J_{s,z}
\end{equation}
to compute resistance and inductance as defined in \cref{eq:R,eq:L}. The vector potential component $A_z$ fully determines the magnetic of a transmission-line segment field since
\begin{subequations}
\begin{align}
B_x &= \frac{1}{i \omega} \partial_y E_z  = \frac{1}{i \omega} \partial_y A_z\,, \\
B_y &= - \frac{1}{i \omega} \partial_y E_z  = - \frac{1}{i \omega} \partial_x A_z\,.
\label{eq:Bfield}
\end{align}
\end{subequations}
We note that the outlined numerical framework can be used to simulate the electromagnetic fields (see \cref{eq:E field,eq:B field}) and transmission-line parameters (see \cref{eq:R,eq:L,eq:C,eq:G}) of arbitrary arrangements of conductor, insulator, and shielding materials.
\section{Results}
\label{sec:Results}
Based on the numerical framework that we introduced in sections~\ref{sec:Formulation} and \ref{sec:Discretization}, we now examine the effect of deformations on transmission-line parameters. We first study the convergence characteristics of the proposed framework by considering a single copper strand and comparing our numerical results with the corresponding analytical theory. After this initial test, we simulate the frequency-dependence of transmission-line parameters for undeformed coaxial geometries and draw a comparison to the deformed case. Finally, we compare our simulations of deformed transmission lines with corresponding experimental data.
\subsection{Single copper strand}
\begin{figure*}
  \centering
  \subfloat[$f = \SI{1}{\hertz}$]{\label{fig:single wire x line}\includegraphics[width=0.5\textwidth]{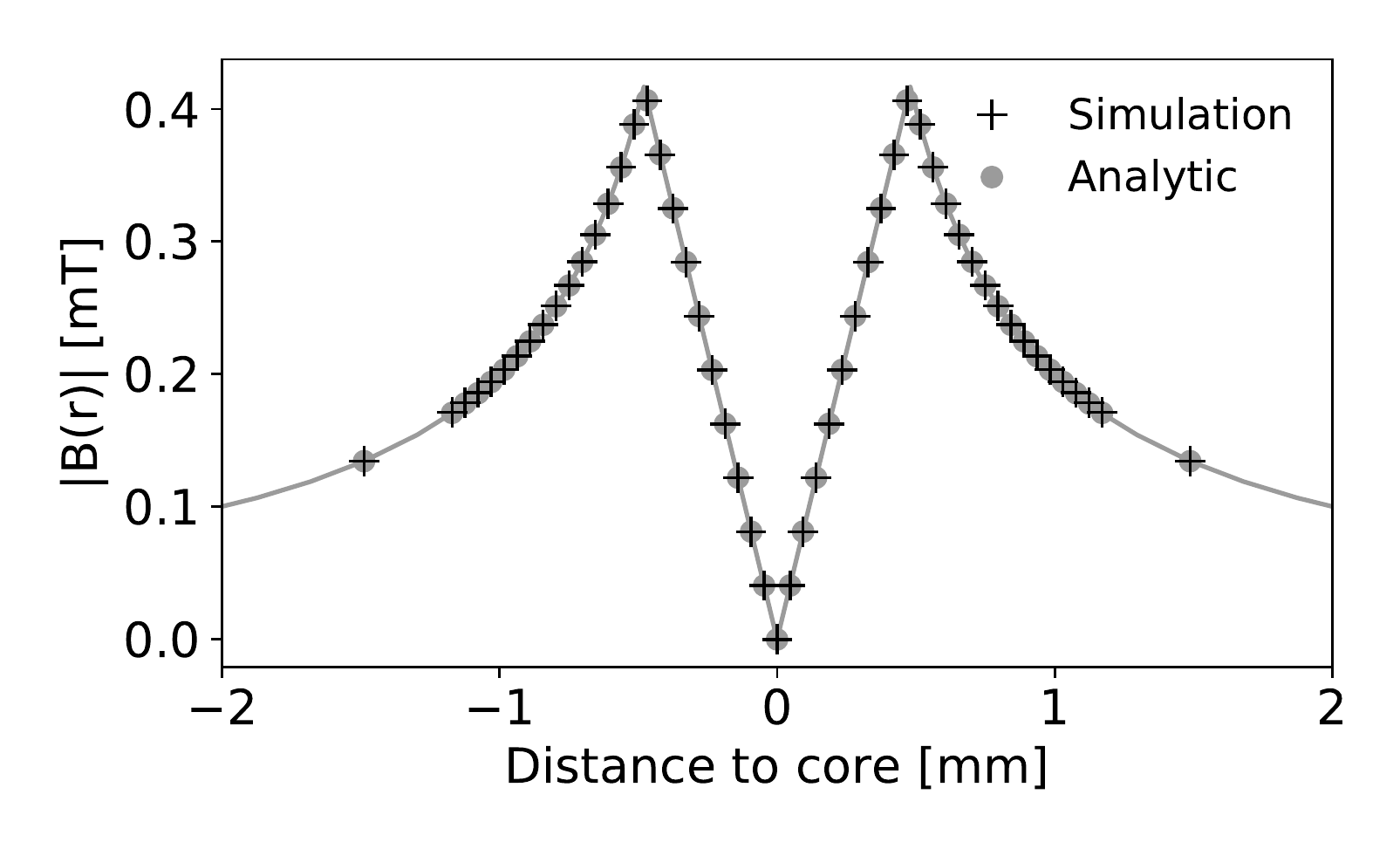}}
  \subfloat[$f = \SI{1}{\hertz}$]{\label{fig:single wire error}\includegraphics[width=0.5\textwidth]{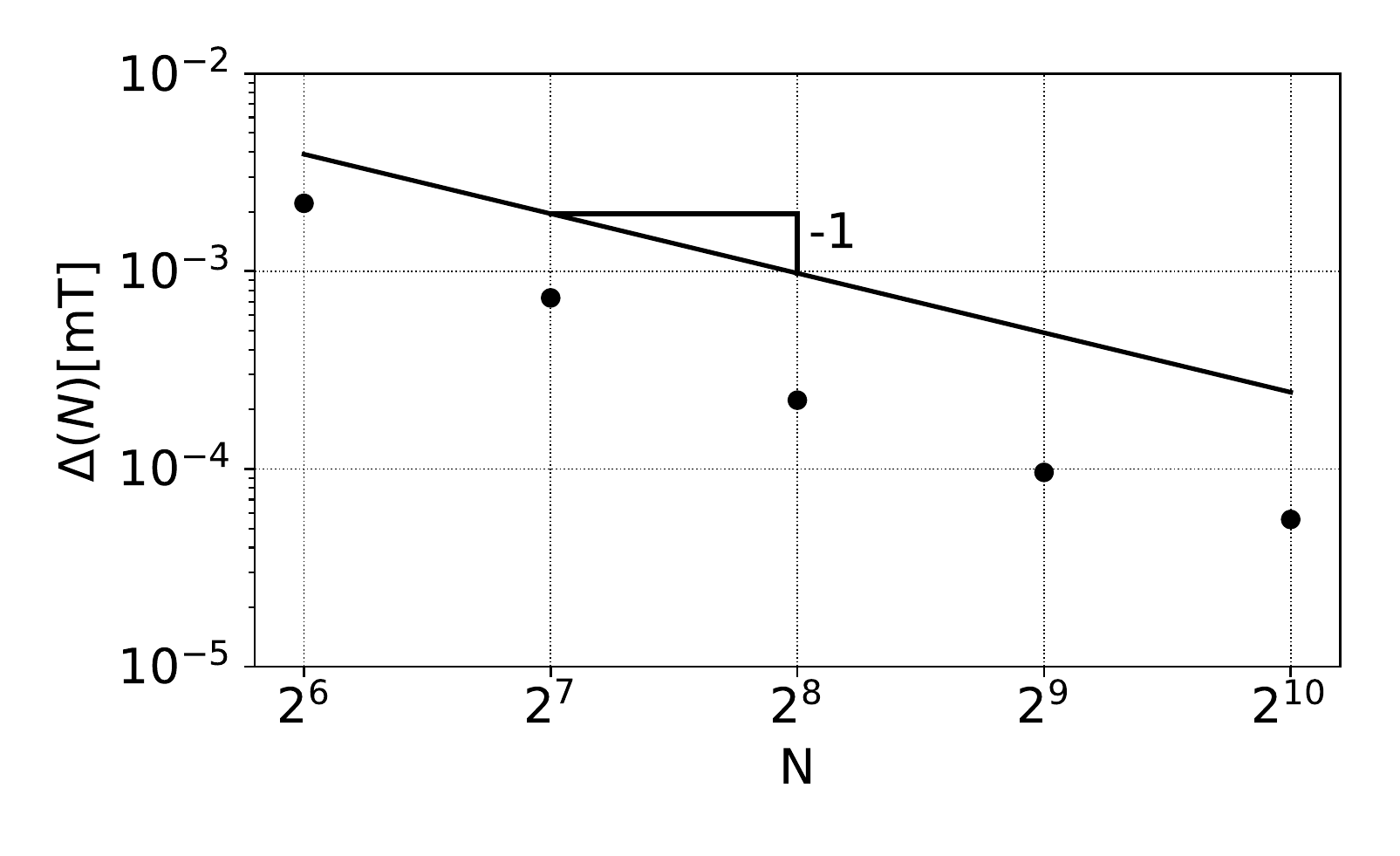}}
  \hfill
  \subfloat[$f = \SI{1}{\mega\hertz}$]{\label{fig:single wire x line high freq}\includegraphics[width=0.5\textwidth]{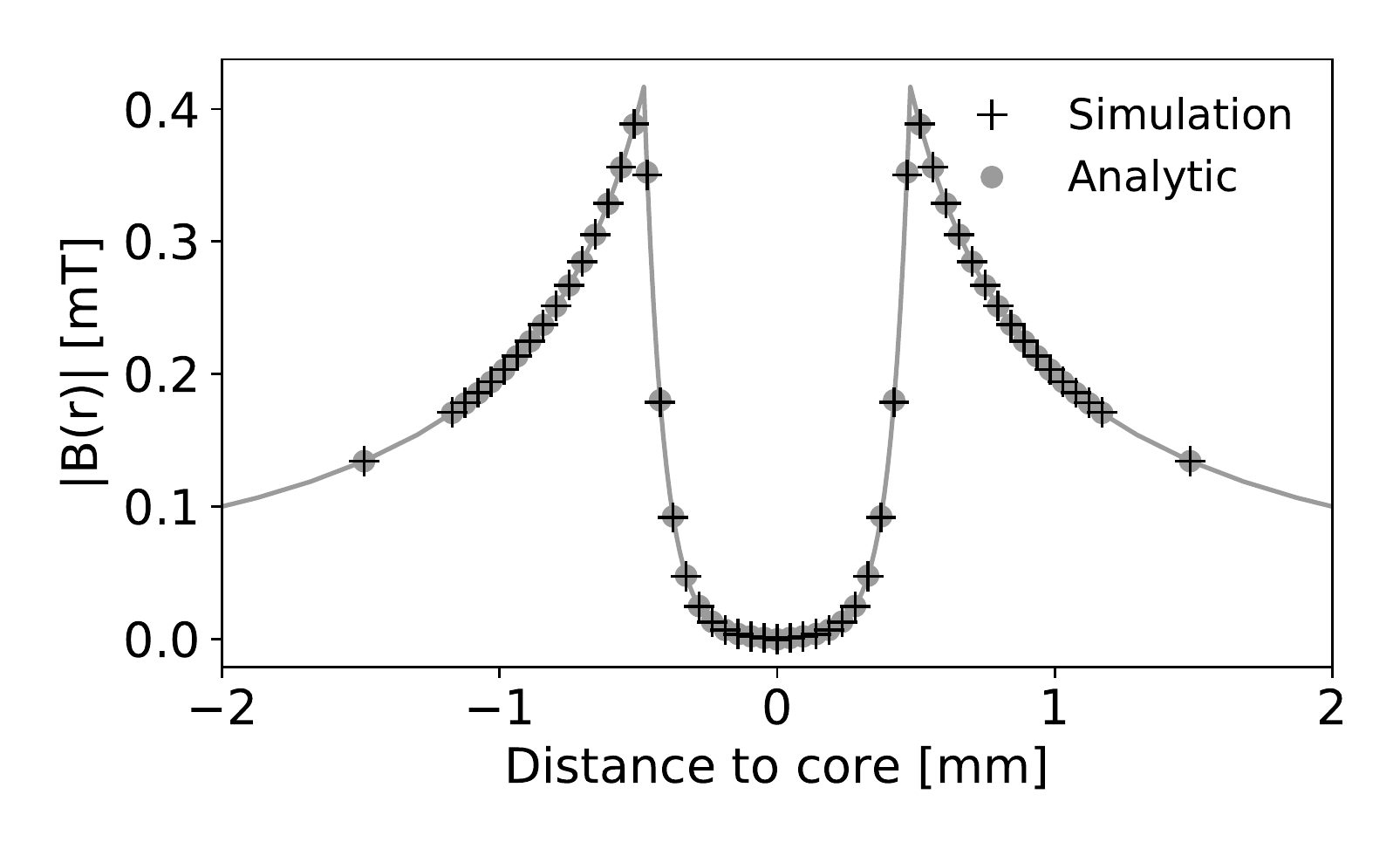}}
  \subfloat[$f = \SI{1}{\mega\hertz}$]{\label{fig:single wire error high freq}\includegraphics[width=0.5\textwidth]{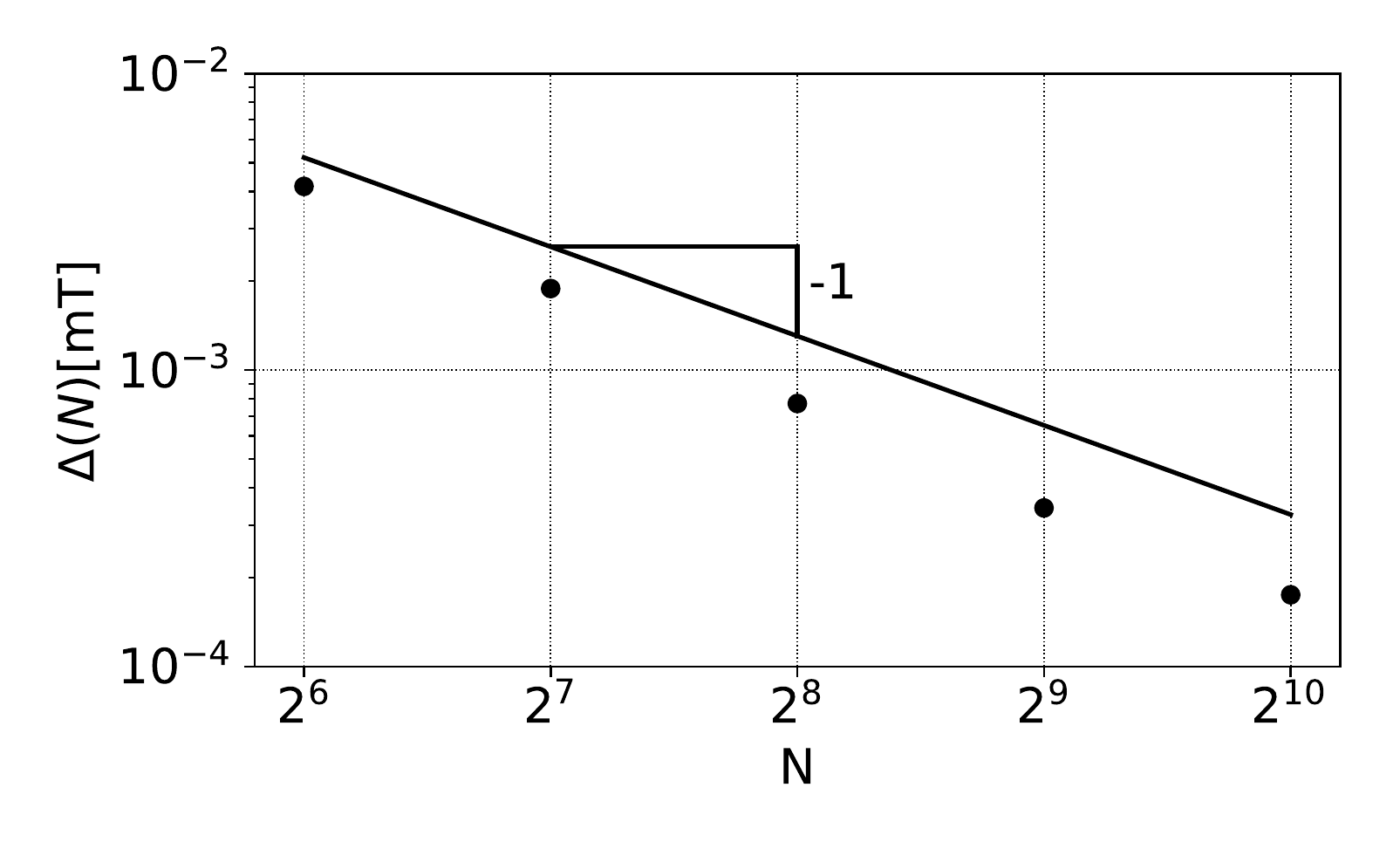}}
  \caption{\textbf{Magnetic field of a single copper strand at low and high frequencies.} In the left panels, we show the absolute value of the magnetic field of a single copper strand surrounded by air for frequencies $f=\SI{1}{\hertz}$ and $f=\SI{1}{\mega\hertz}$. The radius of the strand is $r_1 = \SI{0.48}{\milli\metre} $. The current is set to $I =  \SI{1}{\ampere}$ and the number of grid points along one axis is $N = 256$. The convergence characteristics of our method is shown in the corresponding right panels. The parameters were set to $\epsilon_{\text{r},\text{copper}}^\prime = 1$~\cite{PhotonicCrystals}, $\epsilon_{\text{r},\text{copper}}^{\prime\prime} = 0$, $\sigma_{\text{copper}} = 5.98 \times 10^7 \frac{S}{m}$~\cite{johnson2003high}, $\epsilon_{\text{r},\text{air}}^\prime = 1.00059 $~\cite{DielectricAir}, $\sigma_{\text{air}} = 0$.}
  \label{fig:single wire}
\end{figure*}
To test the proposed numerical framework, we consider a single copper strand of radius $r_1$ and compare the numerically obtained values of the magnetic field with the ones predicted by analytic theory. For low frequencies, the current density within the wire is uniform and using Amp\`ere's law yields
\begin{equation}
|B(r)| = 
\begin{cases}
    \frac{\mu I r}{2 \pi r^2_1}\,, & r\leq r_1\,, \\
    \frac{\mu I}{2 \pi r}\,, & r > r_1\,, \\
\end{cases}
\label{eq:B strain low freq}
\end{equation}
where $r$ is the distance to the center of the strand, $I$ is the current flowing through the strand, and $\mu = \mu_0$. For high frequencies, the current density is not uniformly distributed in the wire, but exhibits a higher density on its surface (\emph{skin effect}). In this case, the electric field is~\cite{Bessel}
\begin{equation}
E_z \left(r\right)= \frac{k I_0\left(k r \right)}{2 \pi \sigma r_1 I_1\left(k r_1 \right)} I\,,
\end{equation}
where $k = \sqrt{i \omega \mu \sigma}$, and $I_0$ and $I_1$ are the modified Bessel functions of first order and first and second kind, respectively. The current density is thus
\begin{equation}
J_z \left(r\right) = \frac{k I_0\left(k r \right)}{2 \pi r_1 I_1\left(k r_1 \right)} I
\label{eq:bessel_J}
\end{equation}
and by applying Amp\`ere's law to \cref{eq:bessel_J} we obtain
\begin{equation}
|B(r)| = 
\begin{cases}
    \frac{\mu I_1\left(k r \right)}{2 \pi r_1 I_1\left(k r_1 \right)} I\,, & r \leq r_1\,, \\
    \frac{\mu I}{2 \pi r}\,, & r > r_1\,. \\
\end{cases}
\label{eq:B strain}
\end{equation}
We now compare our simulation results with the analytical predictions of \cref{eq:B strain} for a single copper strand of radius $r_1=\SI{0.48}{\milli\metre}$ that is surrounded by air. We show the results in figures~\ref{fig:single wire x line} and \ref{fig:single wire x line high freq}, and find good agreement between our simulations and analytical theory. For a frequency of $f=\SI{1}{\mega\hertz}$, we see the influence of the skin effect on the magnetic field. To study the convergence characteristics of our method for different numbers of grid points along one axis $N$, we define the error
\begin{equation}
\Delta\left(N\right) = \frac{1}{M} \sum_{(i,j) \in S}|B_{\text{simulation}}^{i,j}\left(N\right)-B_{\text{analytical}}^{i,j}\left(N\right)|\,,
\end{equation}
where $S=\left\{(i,j)|  r^{i,j} \leq r_c\right\}$ and $r^{i,j}=\sqrt{\left(x^i\right)^2+\left(y^j\right)^2}$. The cut-off radius is denoted by $r_c$, and $M$ is the number of considered points. We use the convention that the value of $r^{i,j}=0$ corresponds to the center of the cooper strand. In our subsequent analysis, we set $r_c= \SI{2}{\milli\metre}$.
The analytical magnetic field values $B_{\text{analytical}}^{i,j}$ correspond to the ones of \cref{eq:B strain}. According to \cref{eq:B field,eq:secondDerivative}, the global error is expected to be of order $\mathcal{O}\left(h\right)$, where $h$ is the distance between the nodes of the grid. We find that the numerically determined error of our method agrees well with the expected scaling (see figures~\ref{fig:single wire error} and \ref{fig:single wire error high freq}).
\subsection{Undeformed coaxial transmission line}
\begin{figure*}
  \centering
  \subfloat[Undeformed coaxial cable]{\label{fig:cross section coax}\includegraphics[width=0.4\textwidth]{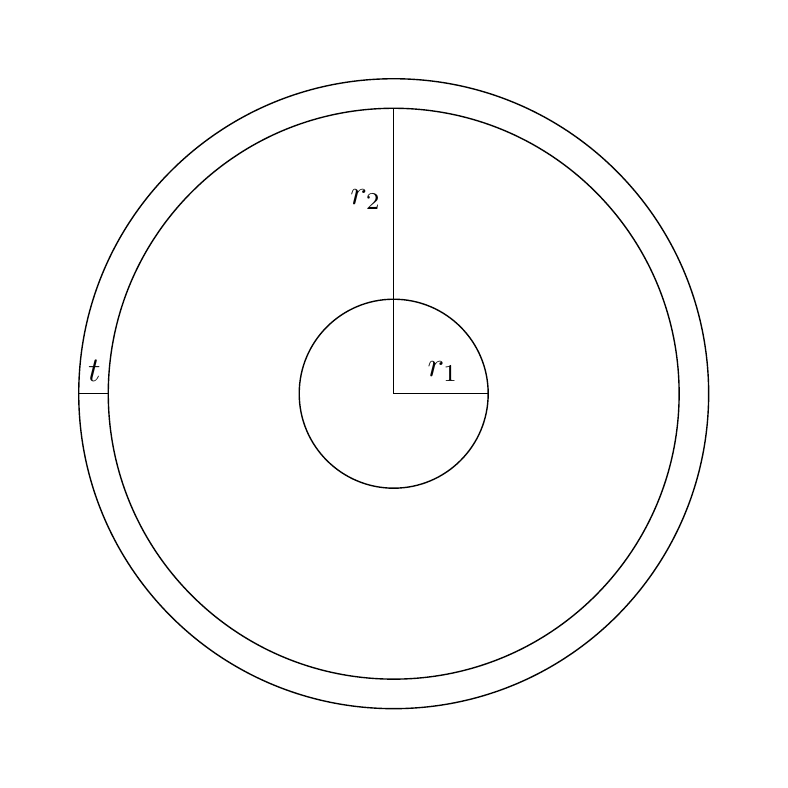}}
  \subfloat[Deformed coaxial cable]{\label{fig:cross section deformed coax}\includegraphics[width=0.4\textwidth]{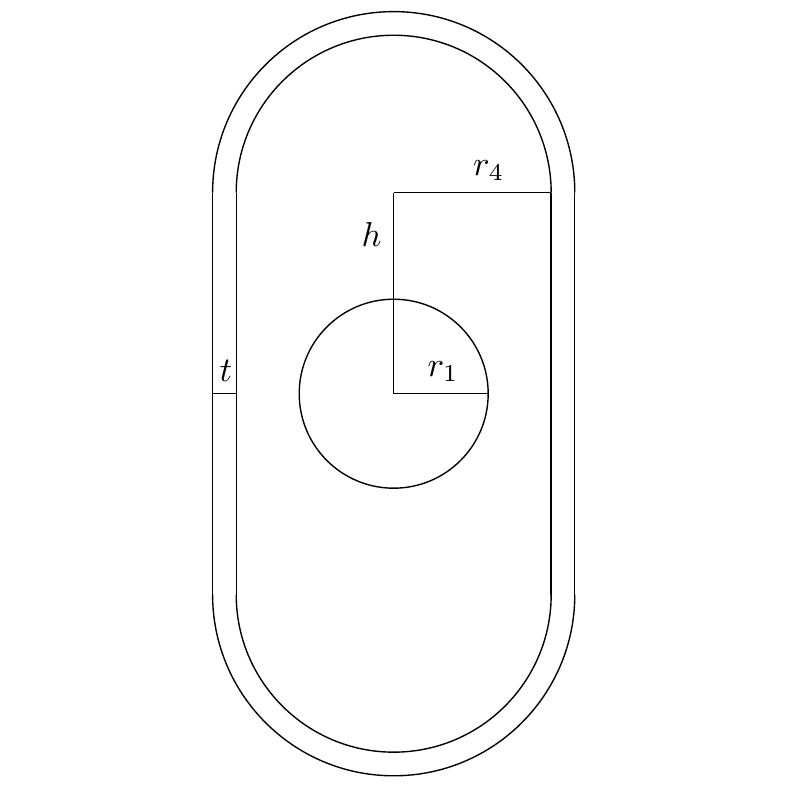}}
  \caption{\textbf{Cross sections of an undeformed and a deformed coaxial cable.} The left panel shows an undeformed coaxial cable with an inner conductor of radius $r_1$. The outer conductor begins at radius $r_2$ away from the center and has thickness $t$. The outer radius of the outer conductor is thus $r_3=r_2+t$. An example of a deformed coaxial cable is shown in the right panel. The radius of the upper and lower half circles is $r_4$. The center of these half circles is located at distance $h$ away from the center of the inner conductor.}
  \label{fig:cross section}
\end{figure*}
\begin{figure*}
  \centering
  \subfloat[$f = \SI{1}{\hertz}$]{\label{fig:coax wire x line}\includegraphics[width=0.5\textwidth]{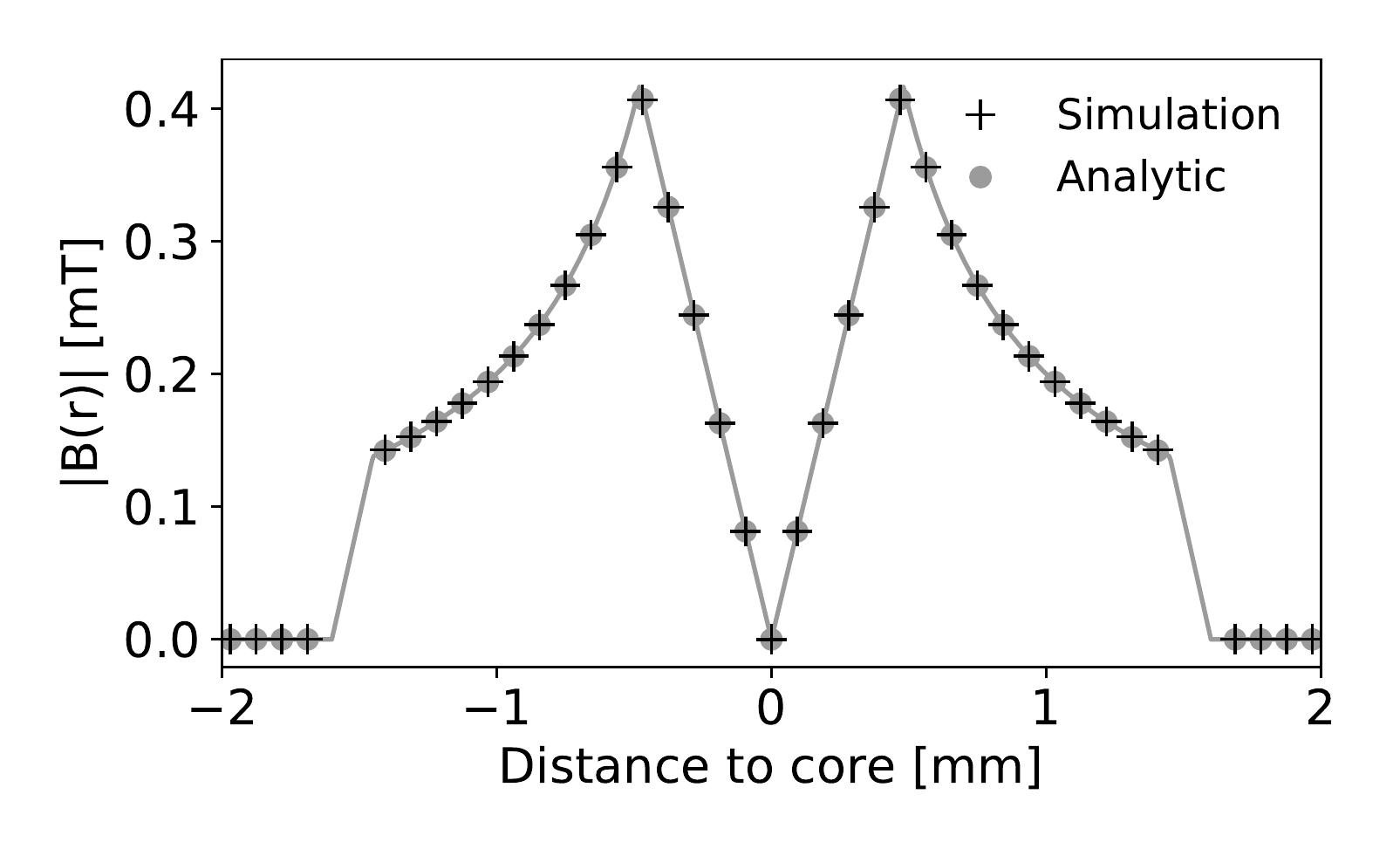}}
  \subfloat[$f = \SI{1}{\hertz}$]{\label{fig:coax wire error}\includegraphics[width=0.5\textwidth]{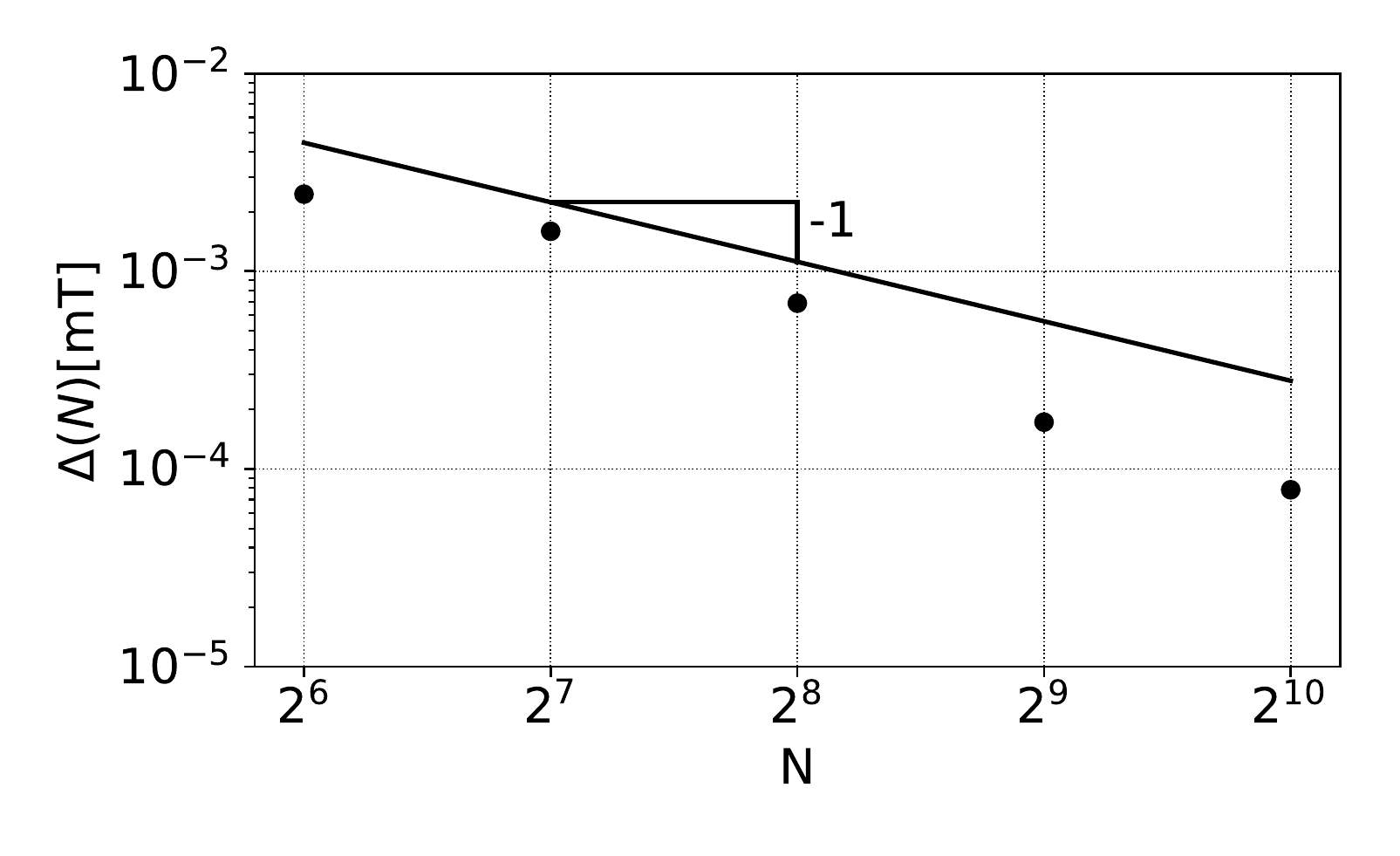}}
  \hfill
  \subfloat[$f = \SI{1}{\mega\hertz}$]{\label{fig:coax x line high freq}\includegraphics[width=0.5\textwidth]{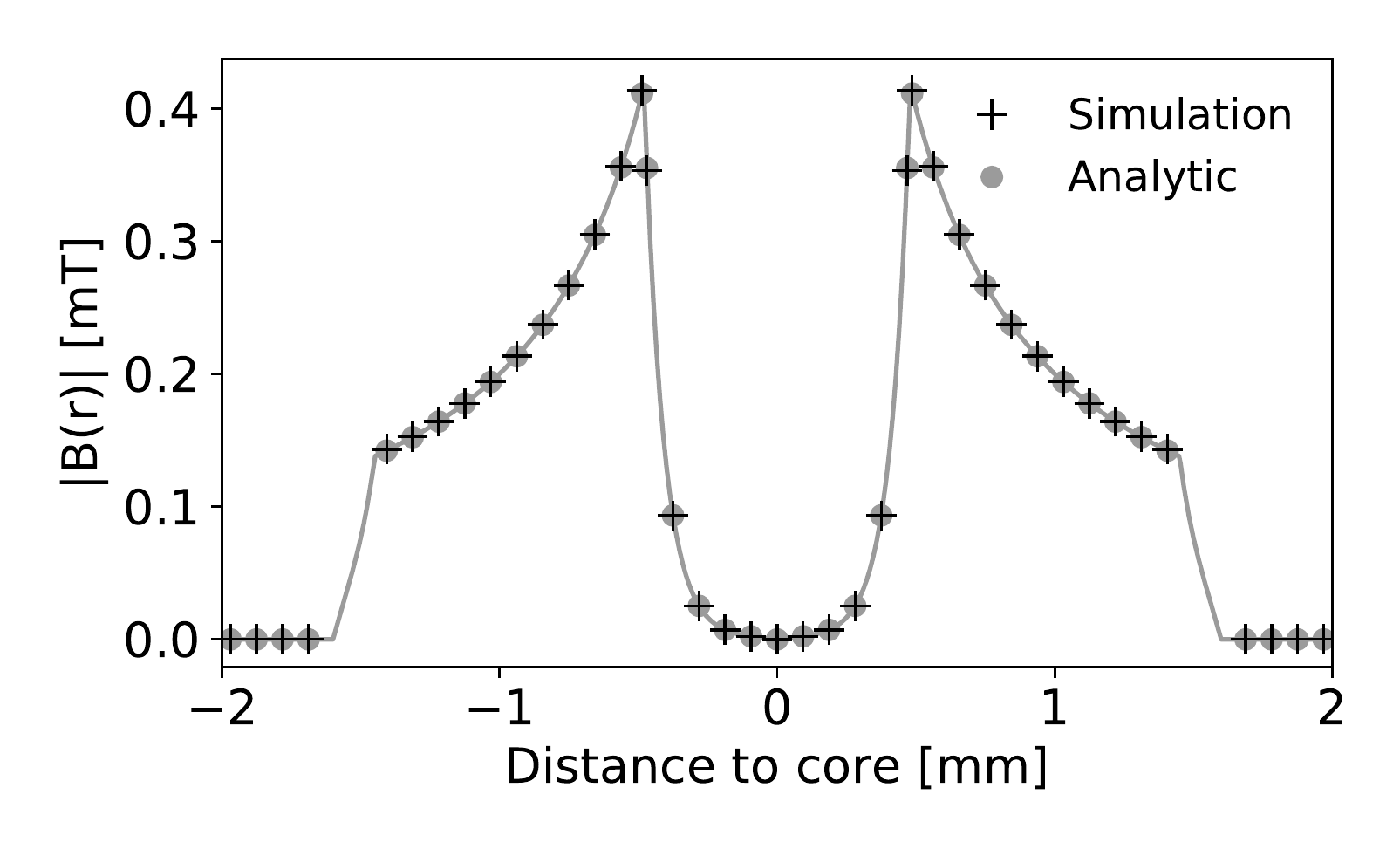}}
  \subfloat[$f = \SI{1}{\mega\hertz}$]{\label{fig:coax error high freq}\includegraphics[width=0.5\textwidth]{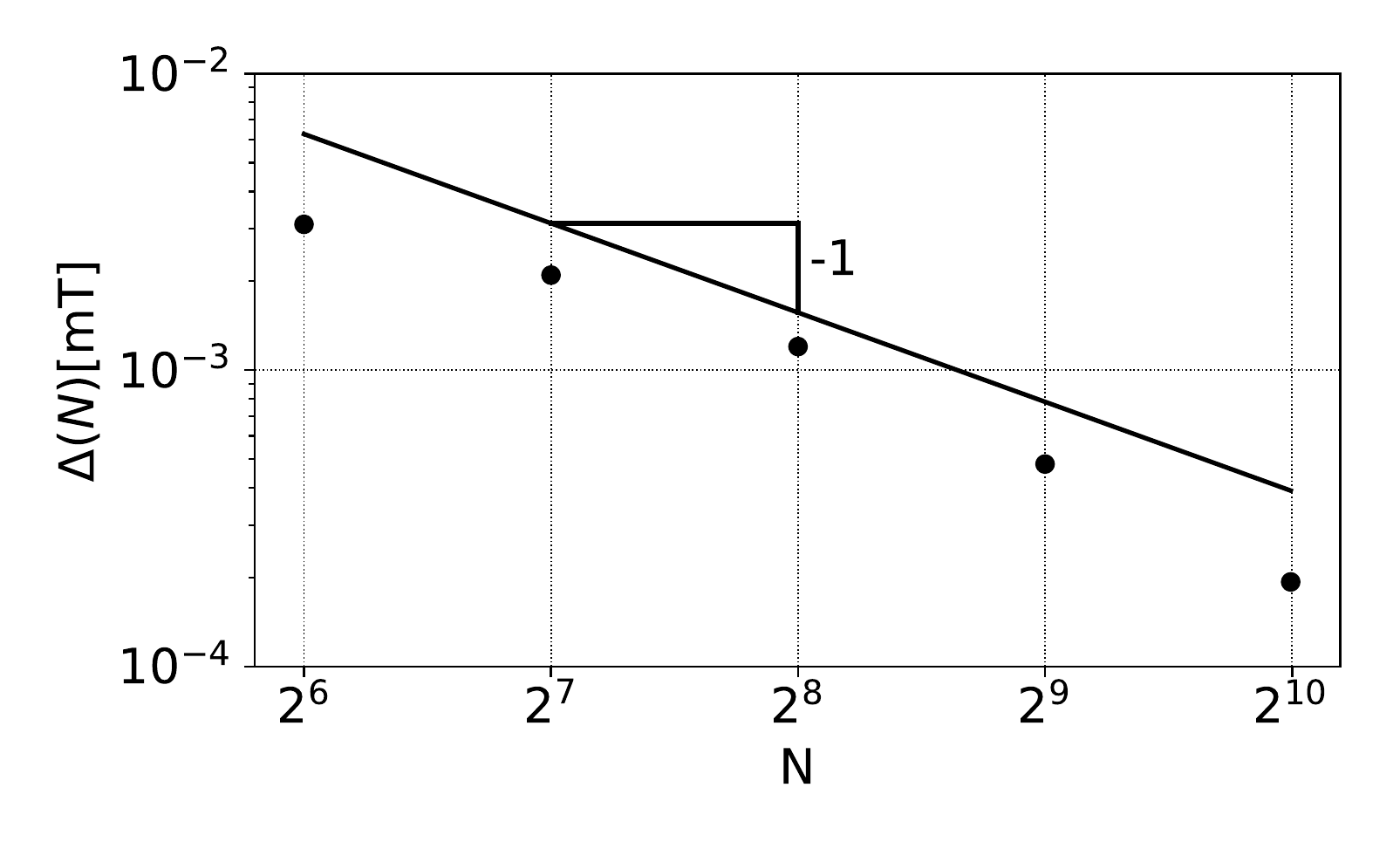}}
  \caption{\textbf{Magnetic field of a coaxial cable at low and high frequencies.} 
In the left panels, we show the absolute value of the magnetic field of a coaxial cable with copper conductors surrounded by air for frequencies $f=\SI{1}{\hertz}$ and $f=\SI{1}{\mega\hertz}$. The current is set to $I =  \SI{1}{\ampere}$ and the current in the inner conductor flows in the opposite direction as the one in the outer conductor. The number of grid points along one axis is $N = 1024$. The convergence characteristics of our method is shown in the corresponding right panels. The remaining parameters are $r_1 =\SI{0.48}{\milli\metre}$, $r_2 = \SI{1.45}{\milli\metre}$, $r_3 = \SI{1.6}{\milli\metre}$, $\epsilon_{\text{r},\text{dielectric}}^\prime = 2.25$, $\sigma_{\text{dielectric}}=0$, $\tan\left(\delta\right) = 10^{-3}$, $\epsilon_{\text{r},\text{copper}}^\prime = 1$~\cite{PhotonicCrystals}, $\sigma_{\text{copper}} = 5.98 \times 10^7 \frac{S}{m}$~\cite{johnson2003high}, $\epsilon_{\text{r},\text{air}}^\prime = 1.00059$~\cite{DielectricAir}, $\sigma_{\text{air}} = 0$.}
  \label{fig:coax}
\end{figure*}
To examine the effect of deformations on transmission-line parameters, we first simulate an undeformed coaxial cable (see figure~\ref{fig:cross section coax}). In this way, we can later compare our results on deformed transmission lines with those obtained for an undeformed reference. We denote the radius of the inner conductor by $r_1$ and $r_2$ is the inner radius of the outer conductor of thickness $t$. Therefore, the outer radius of the outer conductor is $r_3=r_2+t$. Similarly to the magnetic field of a single copper strand (see \cref{eq:B strain}), we find for an undeformed coaxial transmission line
\begin{equation}
|B(r)| = 
\begin{cases}
    \frac{\mu I_1\left(k r \right)}{2 \pi r_1 I_1\left(k r_1 \right)} I\,, & r \leq r_1\,, \\
    \frac{\mu I}{2 \pi r}\,, & r_1\leq r \leq r_2\,, \\
    \frac{\mu I_1\left(k \left(r_3-r\right) \right)}{2 \pi r_2  I_1\left(k \left(r_3-r_2\right) \right)} I\,, & r_2\leq r \leq r_3\,, \\
    0 \,,& r >r_3\,. \\
\end{cases}
\label{eq:B coax}
\end{equation}
The expressions for $r\leq r_2$ and $r\geq r_3$ are exact. The magnetic field vanishes for $r\geq r_3$, because the currents in the inner and outer conductors are oriented in the opposite directions. For the magnetic field in the outer conductor, we use an approximation that assumes that the current density distribution is the same as for a wire of radius $t$. In figures~\ref{fig:coax wire x line} and \ref{fig:coax x line high freq}, we see that the simulated magnetic field values agree well with the predictions of \cref{eq:B coax}. For large enough frequencies, we recognize the influence of the skin effect (see figure~\ref{fig:coax x line high freq}). As for the single copper strand, we examine the convergence properties of our method for the coaxial geometry (see figures~\ref{fig:coax wire error} and \ref{fig:coax error high freq}). Due to the skin effect, a larger number of grid cells is required for higher frequencies than for lower ones to resolve the electromagnetic fields.
\begin{figure*}[h]
  \centering
  \subfloat{\label{fig:coax resistance}\includegraphics[width=0.5\textwidth]{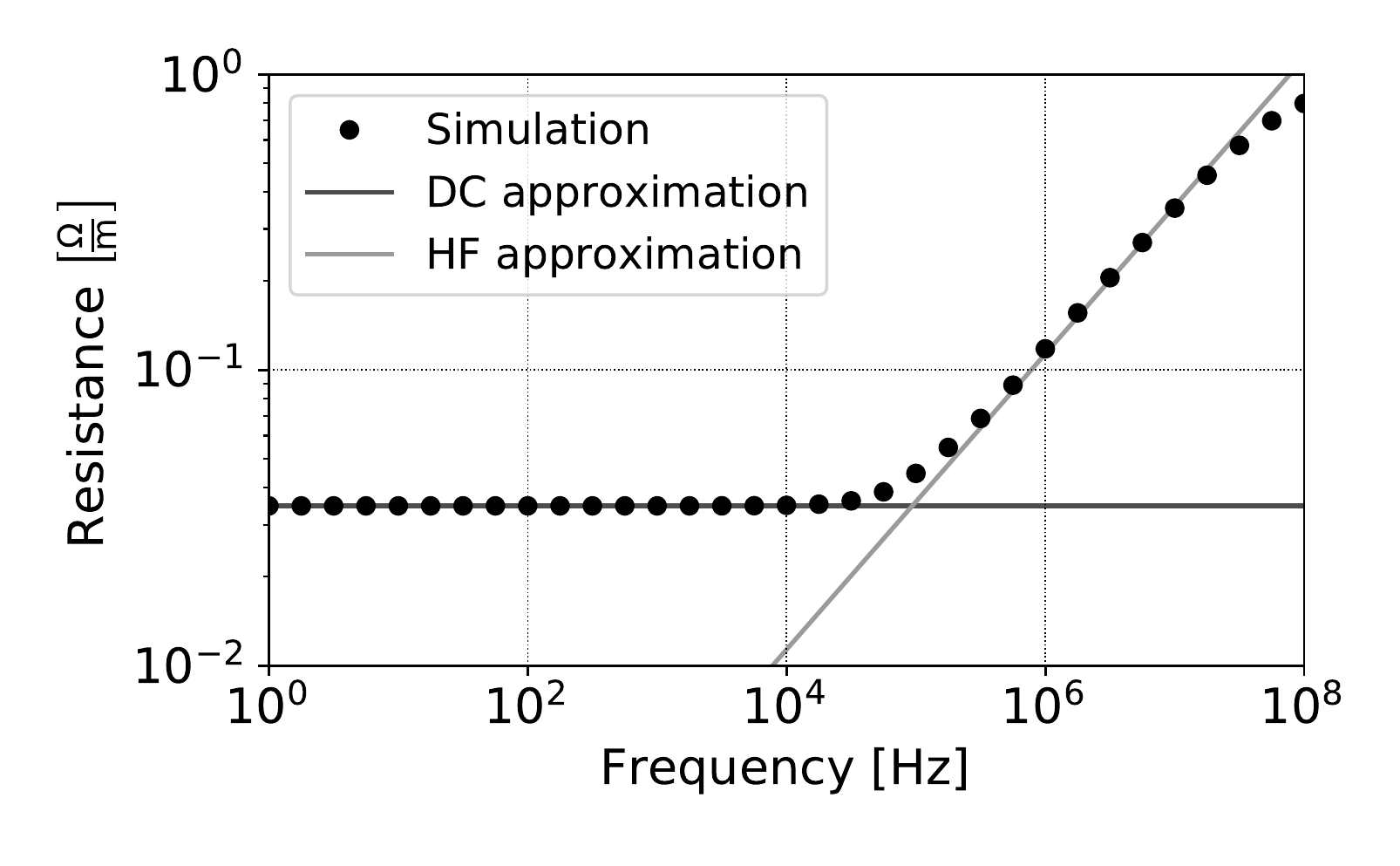}}
  \subfloat{\label{fig:coax inductance}\includegraphics[width=0.5\textwidth]{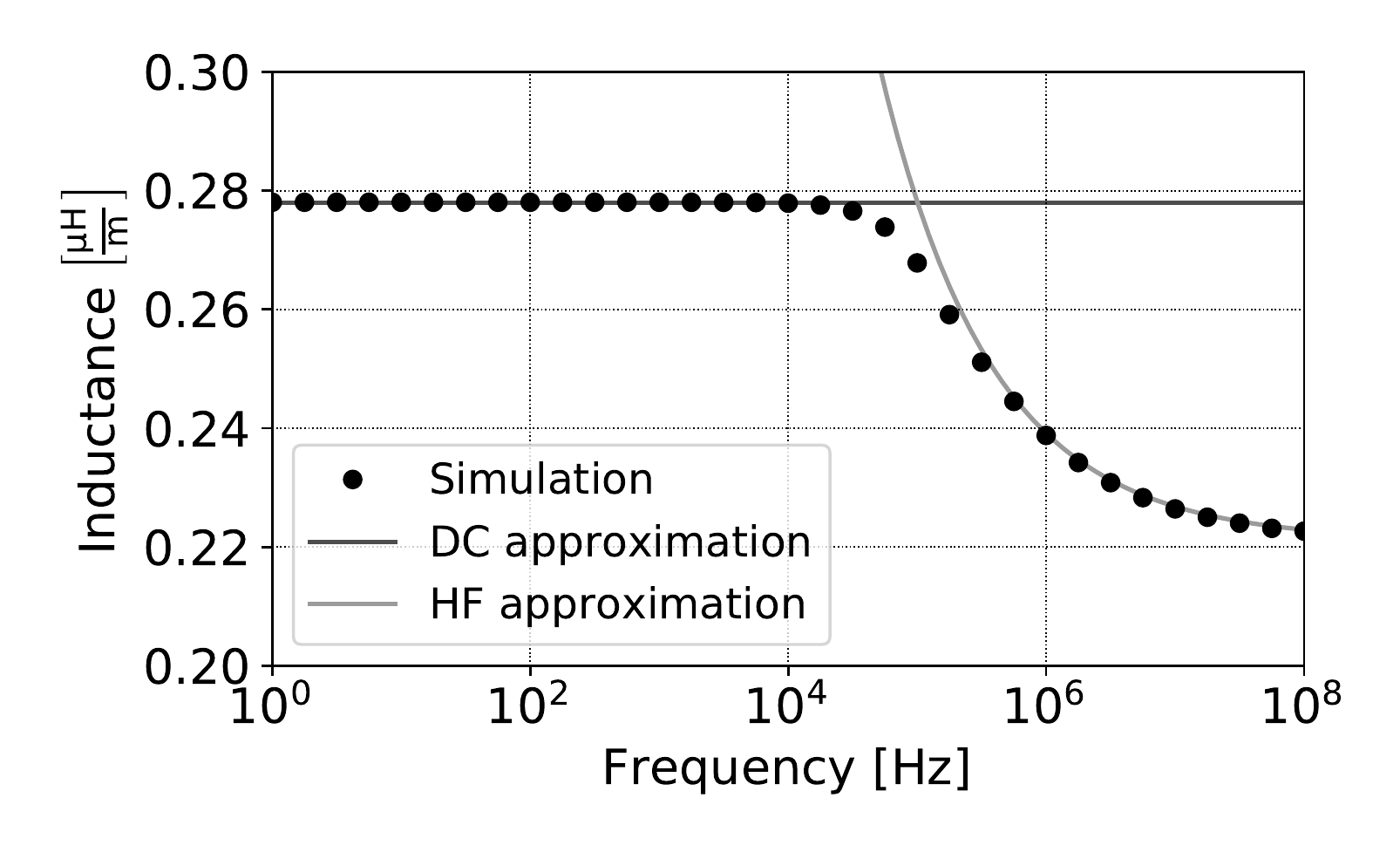}}
  \hfill
  \subfloat{\label{fig:coax capacity}\includegraphics[width=0.5\textwidth]{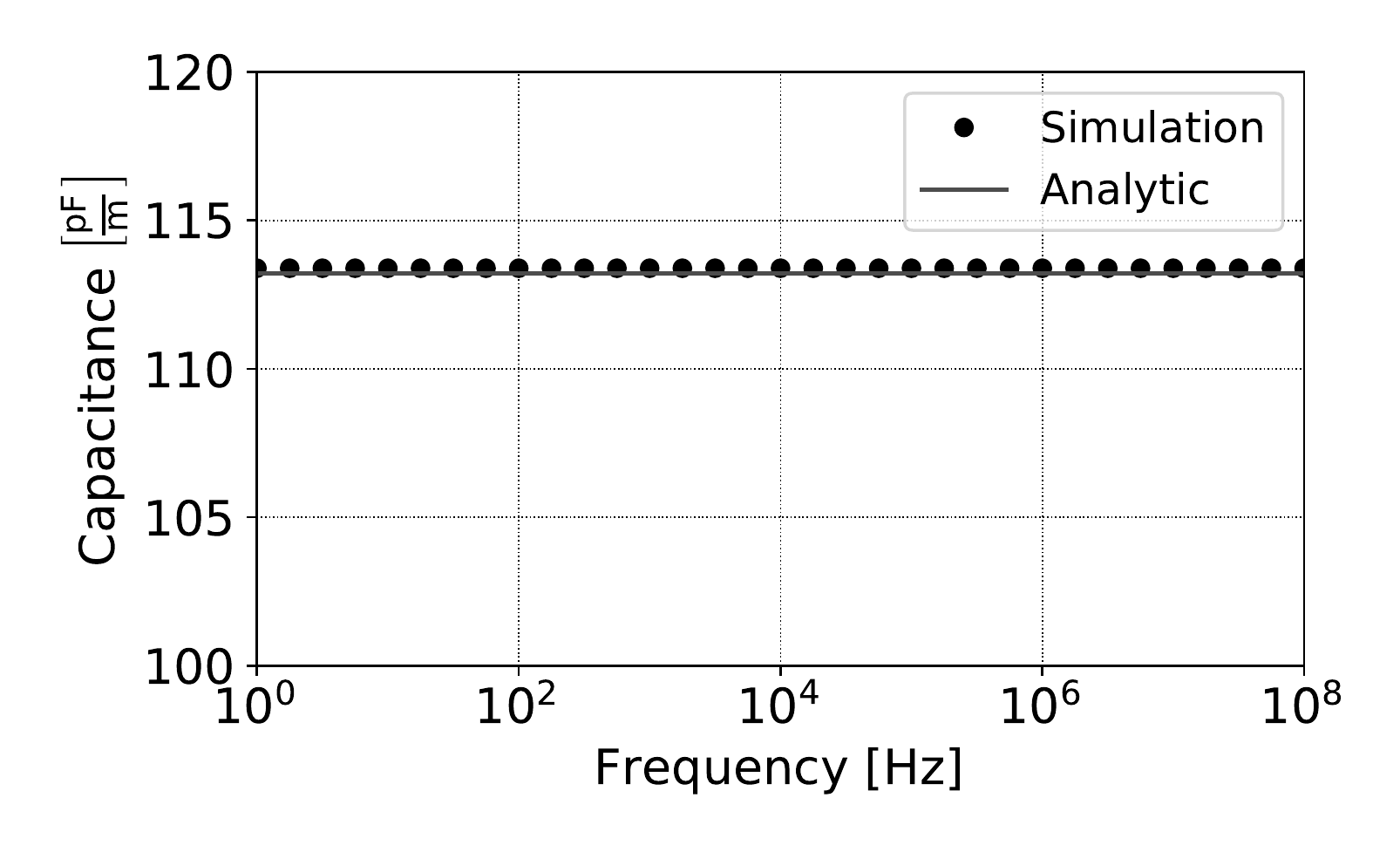}}
  \subfloat{\label{fig:coax conductance}\includegraphics[width=0.5\textwidth]{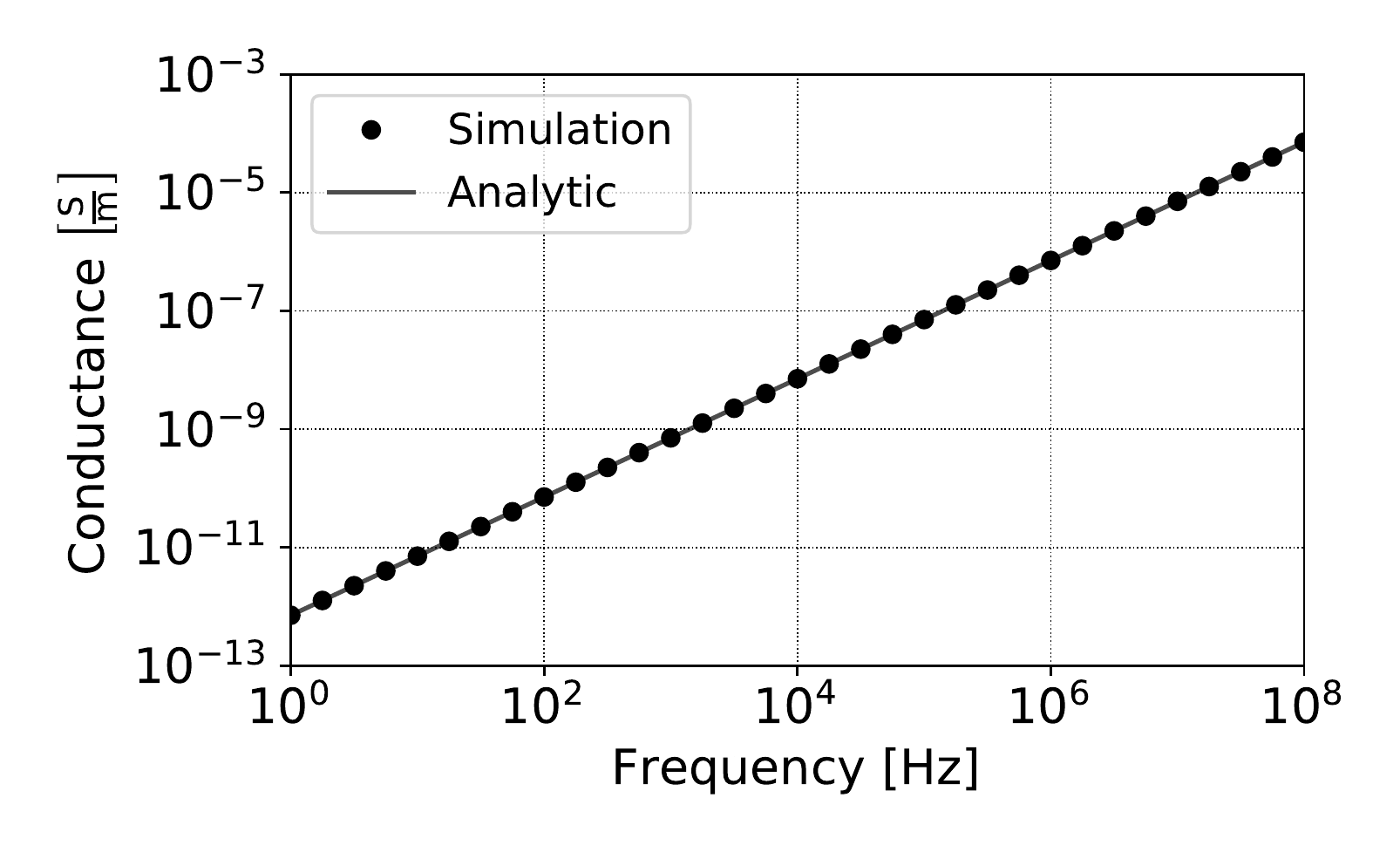}}
  \hfill
  \subfloat{\includegraphics[width=0.5\textwidth]{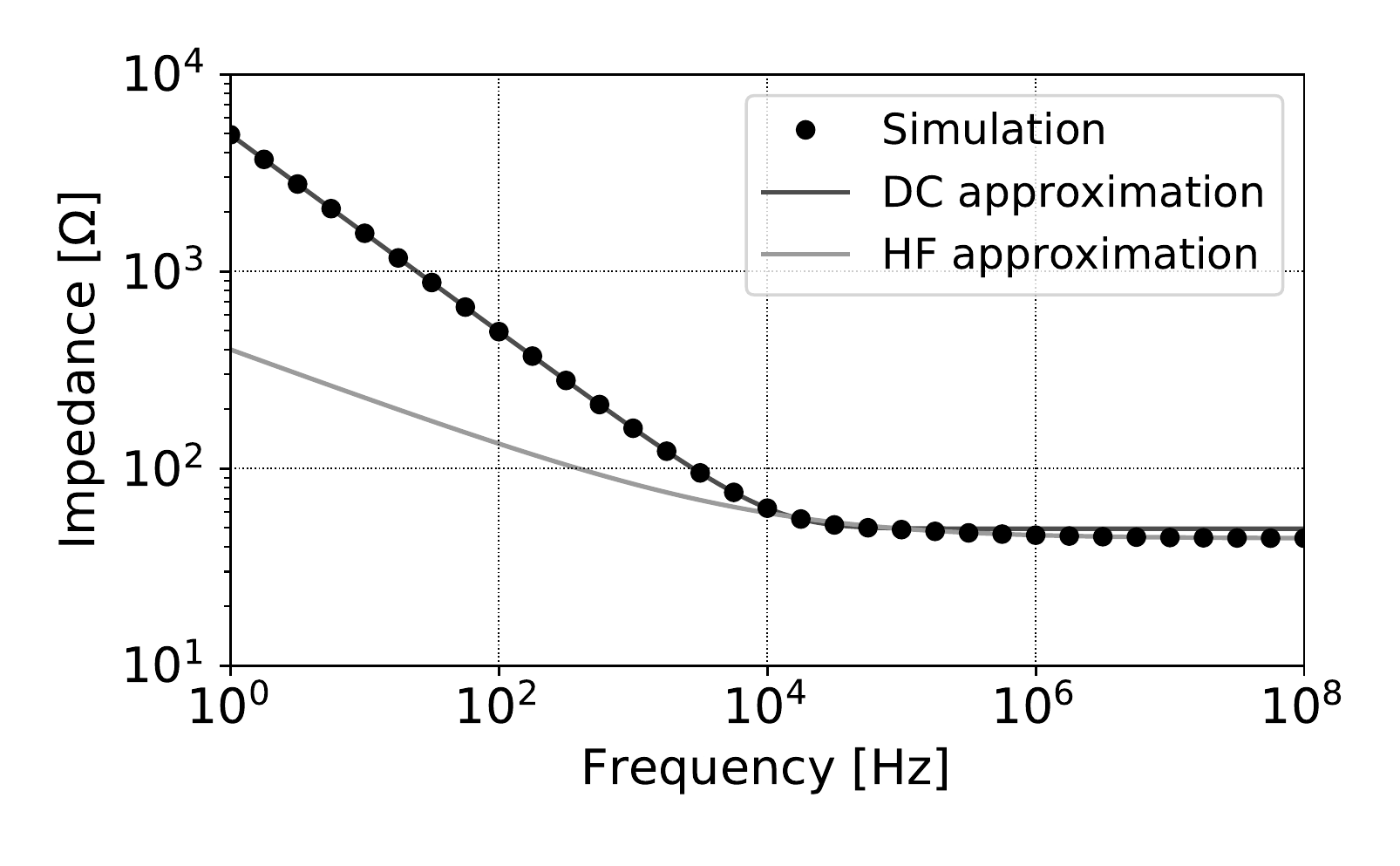}}
  \subfloat{\includegraphics[width=0.5\textwidth]{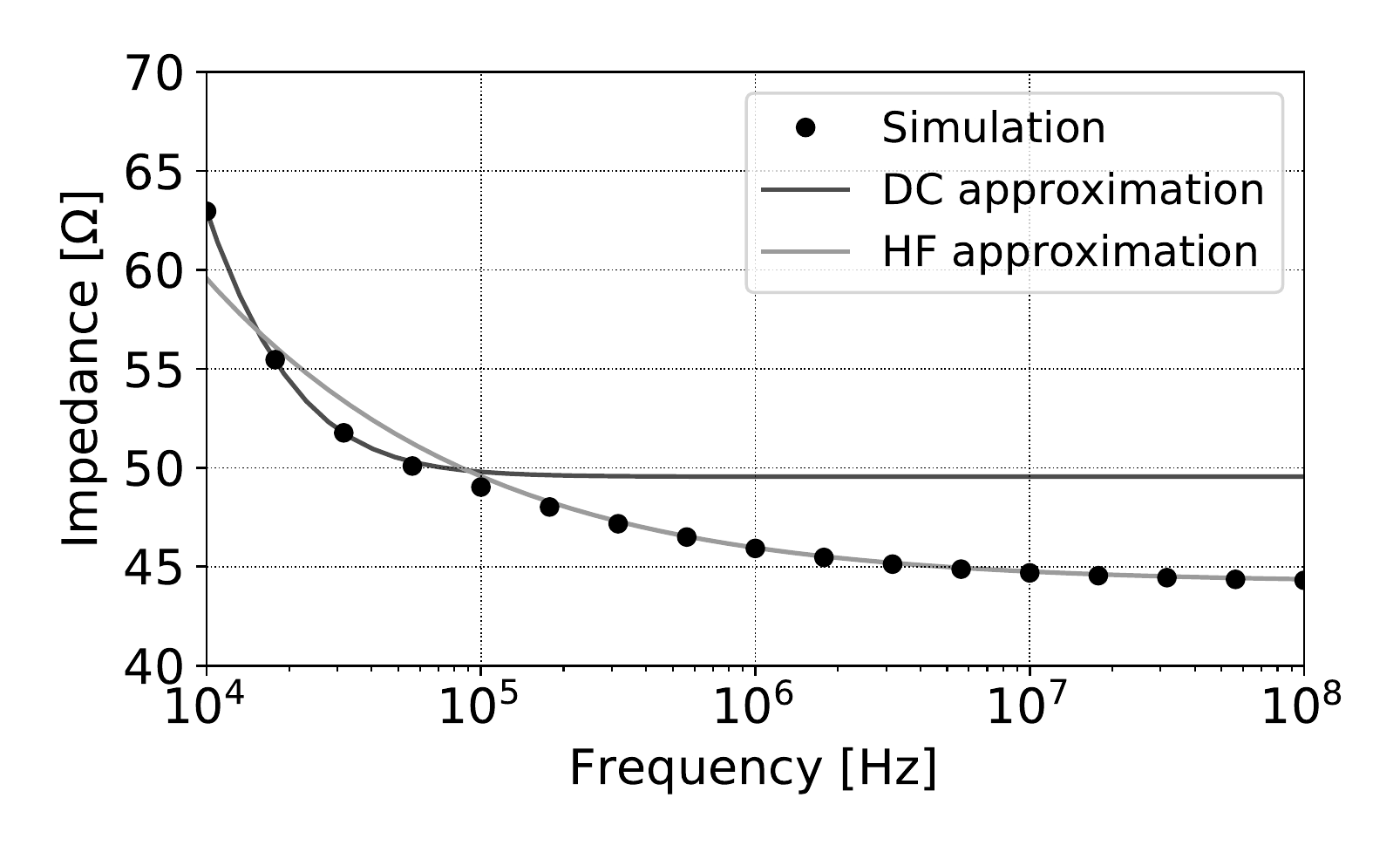}}
  \caption{\textbf{Transmission-line parameters of an undeformed coaxial cable.} We show the transmission-line parameters and the characteristic impedance (simulation and analytic approximations) as defined by~\cref{eq:R,eq:L,eq:C,eq:G,eq:impedance} for an undeformed coaxial cable. The number of grid points along one axis is $N= 512$  and the remaining parameters are $r_1 =\SI{0.48}{\milli\metre}$, $r_2 = \SI{1.45}{\milli\metre}$, $r_3 = \SI{1.6}{\milli\metre}$, $\epsilon_{\text{r},\text{dielectric}}^\prime = 2.25$, $\sigma_{\text{dielectric}}=0$, $\tan\left(\delta\right) = 10^{-3}$, $\epsilon_{\text{r},\text{copper}}^\prime = 1$~\cite{PhotonicCrystals}, $\sigma_{\text{copper}} = 5.98 \times 10^7 \frac{S}{m}$~\cite{johnson2003high}, $\epsilon_{\text{r},\text{air}}^\prime = 1.00059$~\cite{DielectricAir}, $\sigma_{\text{air}} = 0$.}
  \label{fig:coax2}
\end{figure*}
\begin{figure*}[hbp!]
  \centering
  \subfloat{\includegraphics[width=0.5\textwidth]{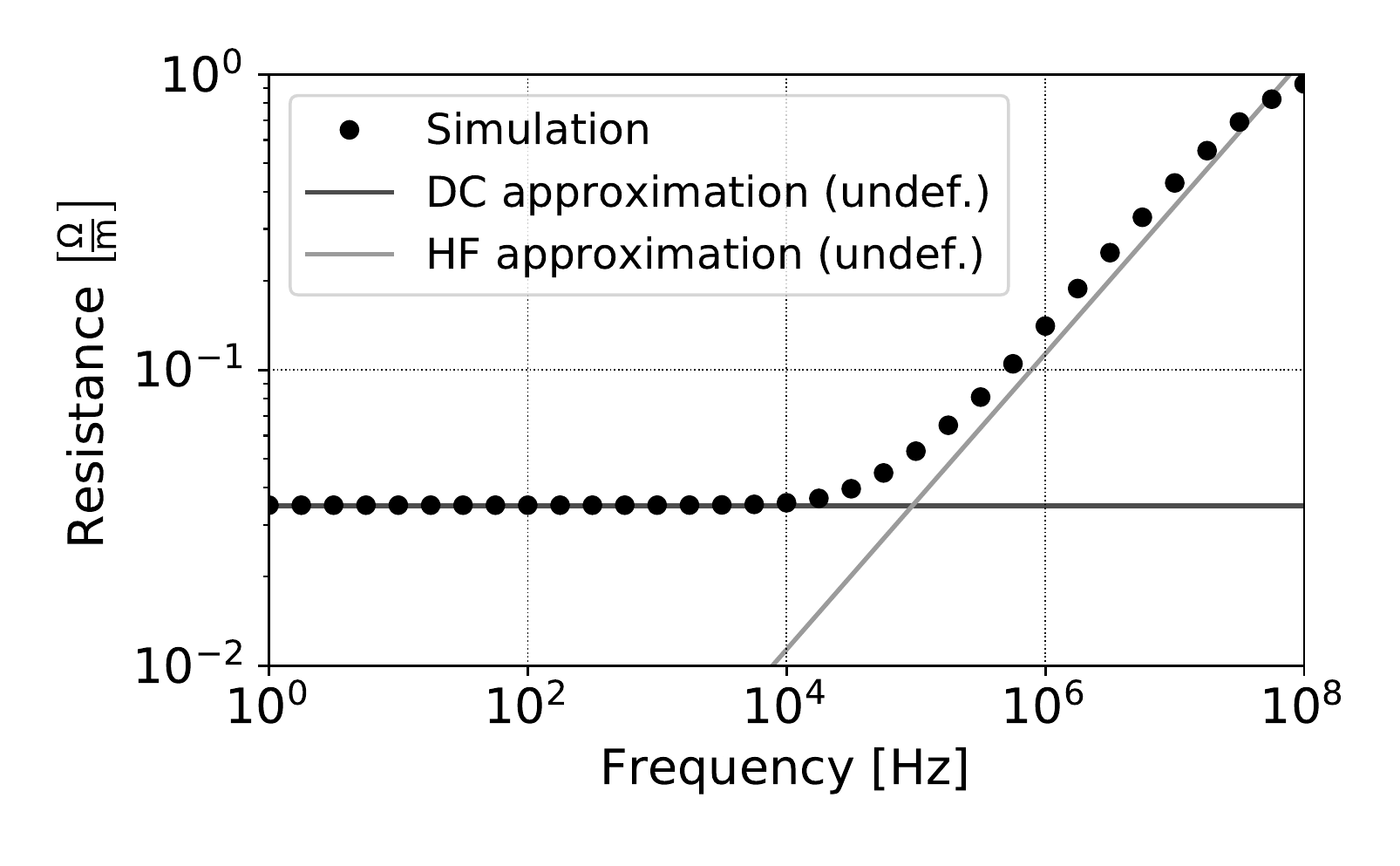}}
  \subfloat{\includegraphics[width=0.5\textwidth]{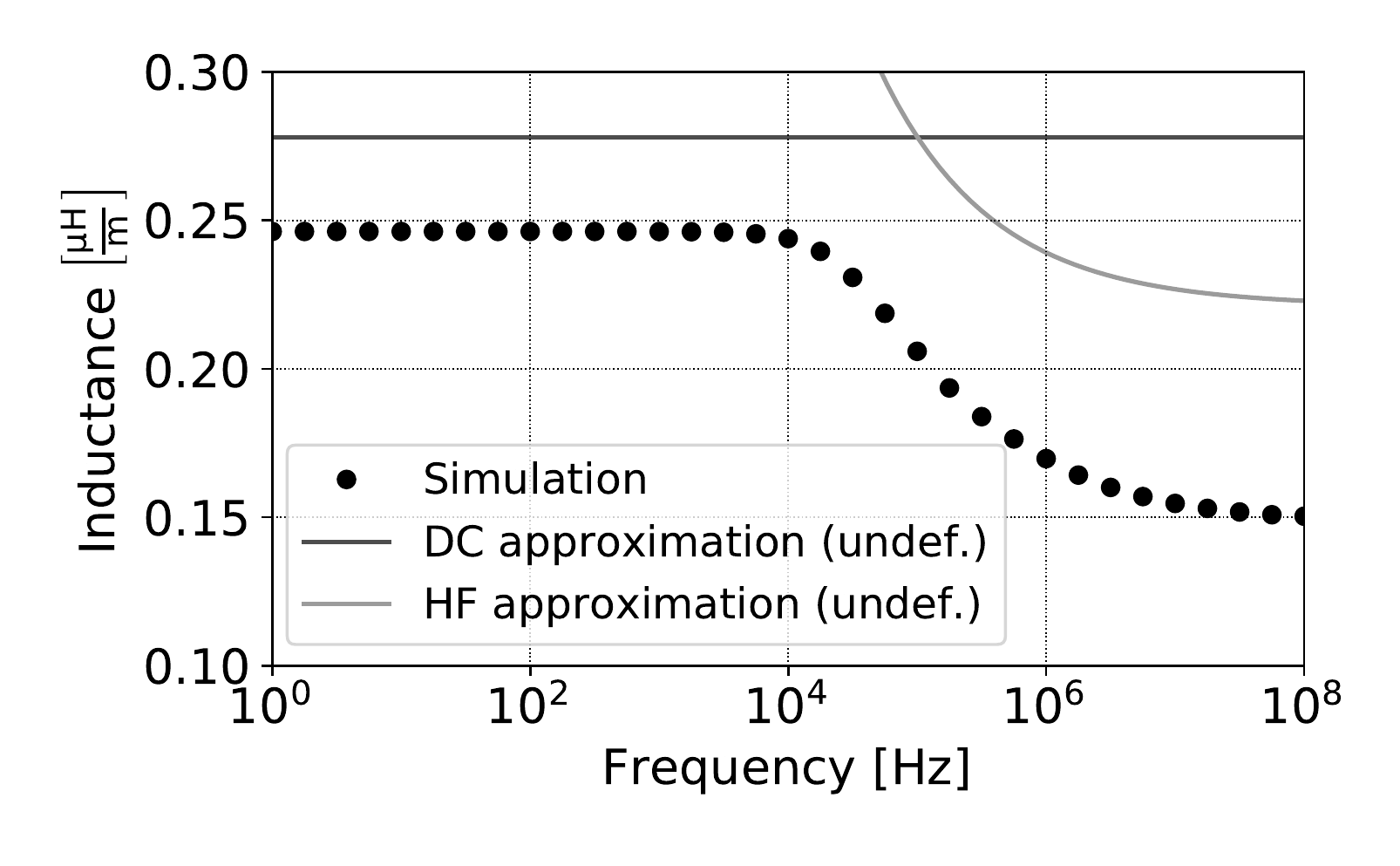}}
  \hfill
  \subfloat{\includegraphics[width=0.5\textwidth]{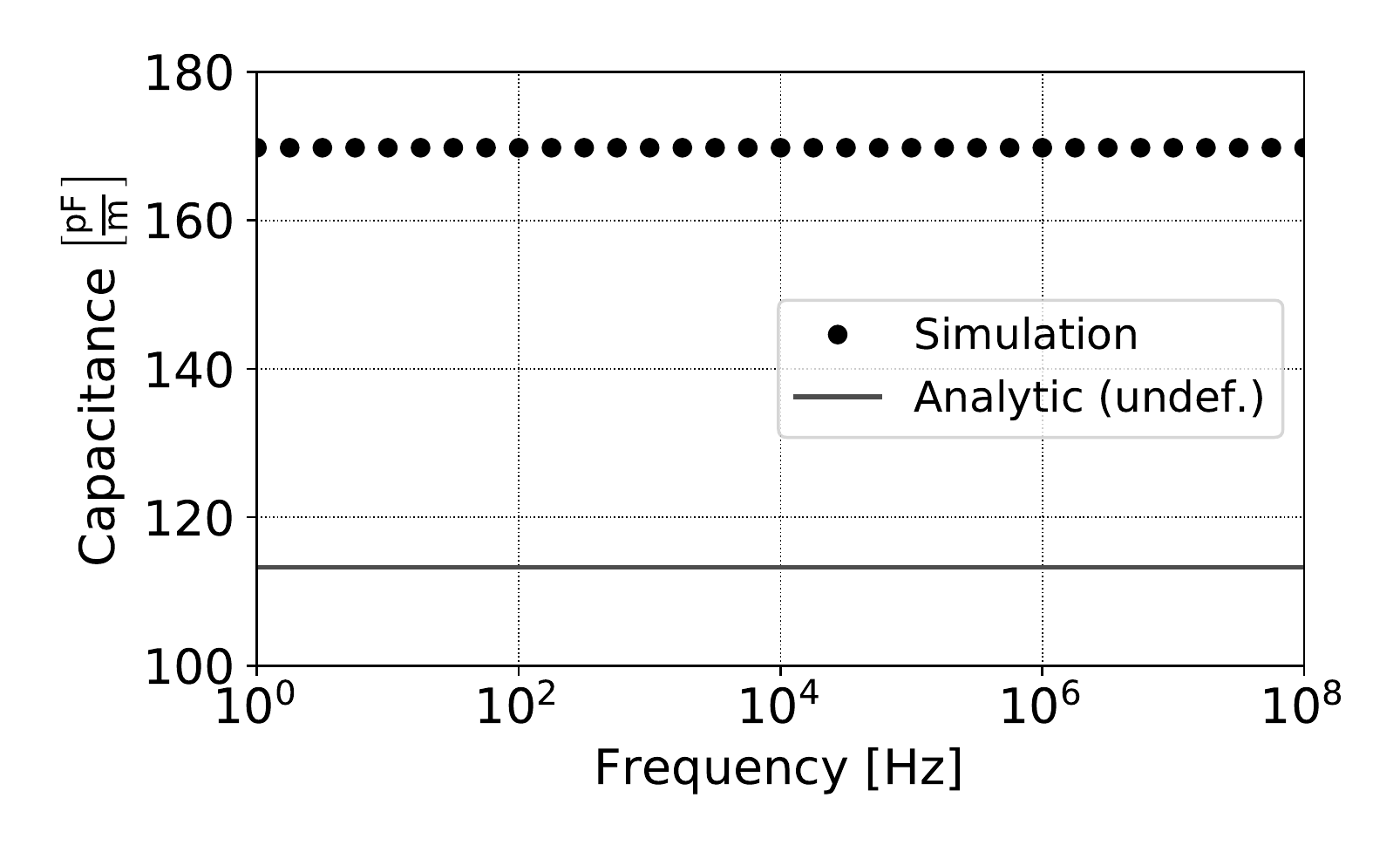}}
  \subfloat{\includegraphics[width=0.5\textwidth]{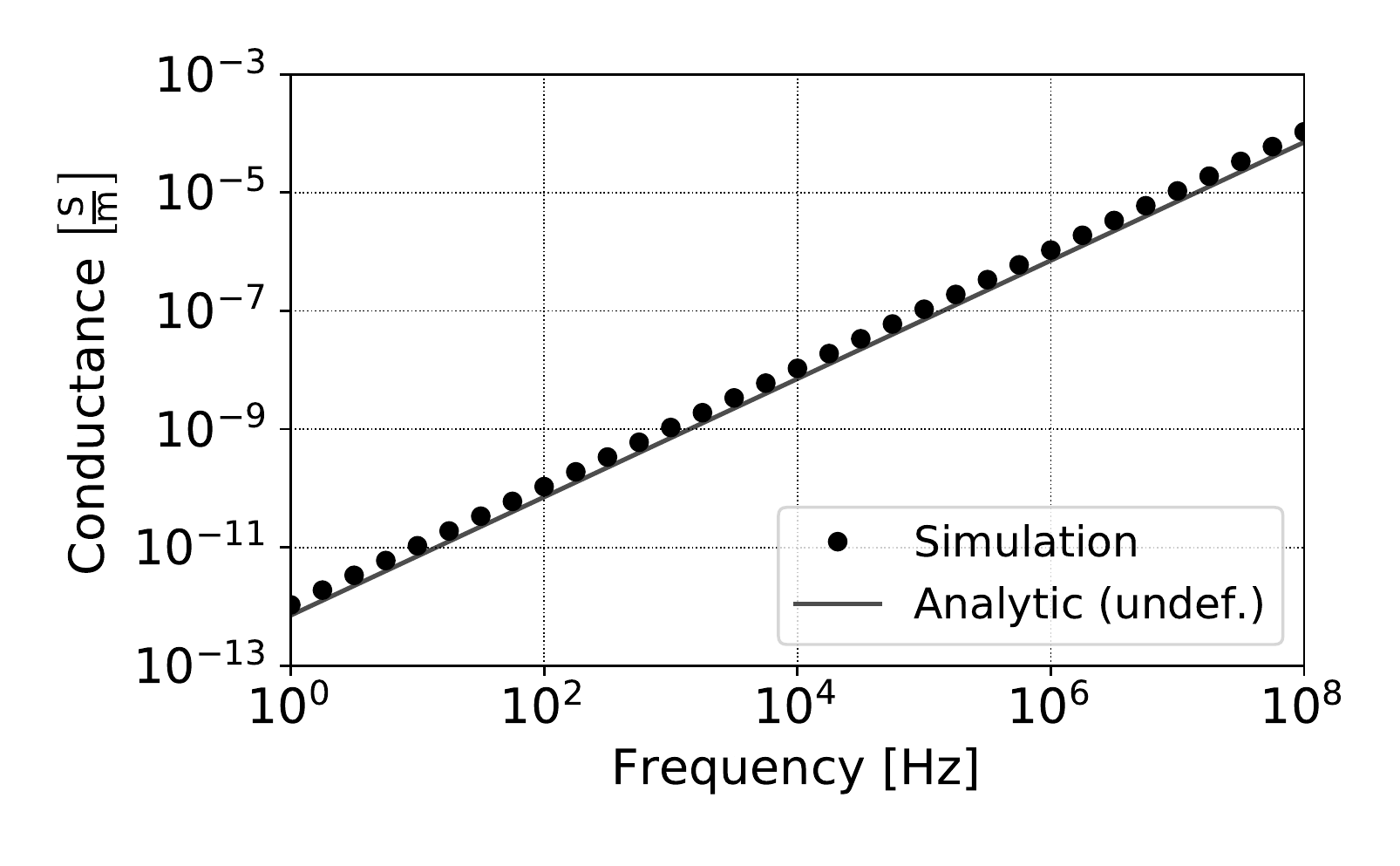}}
  \hfill
  \subfloat{\includegraphics[width=0.5\textwidth]{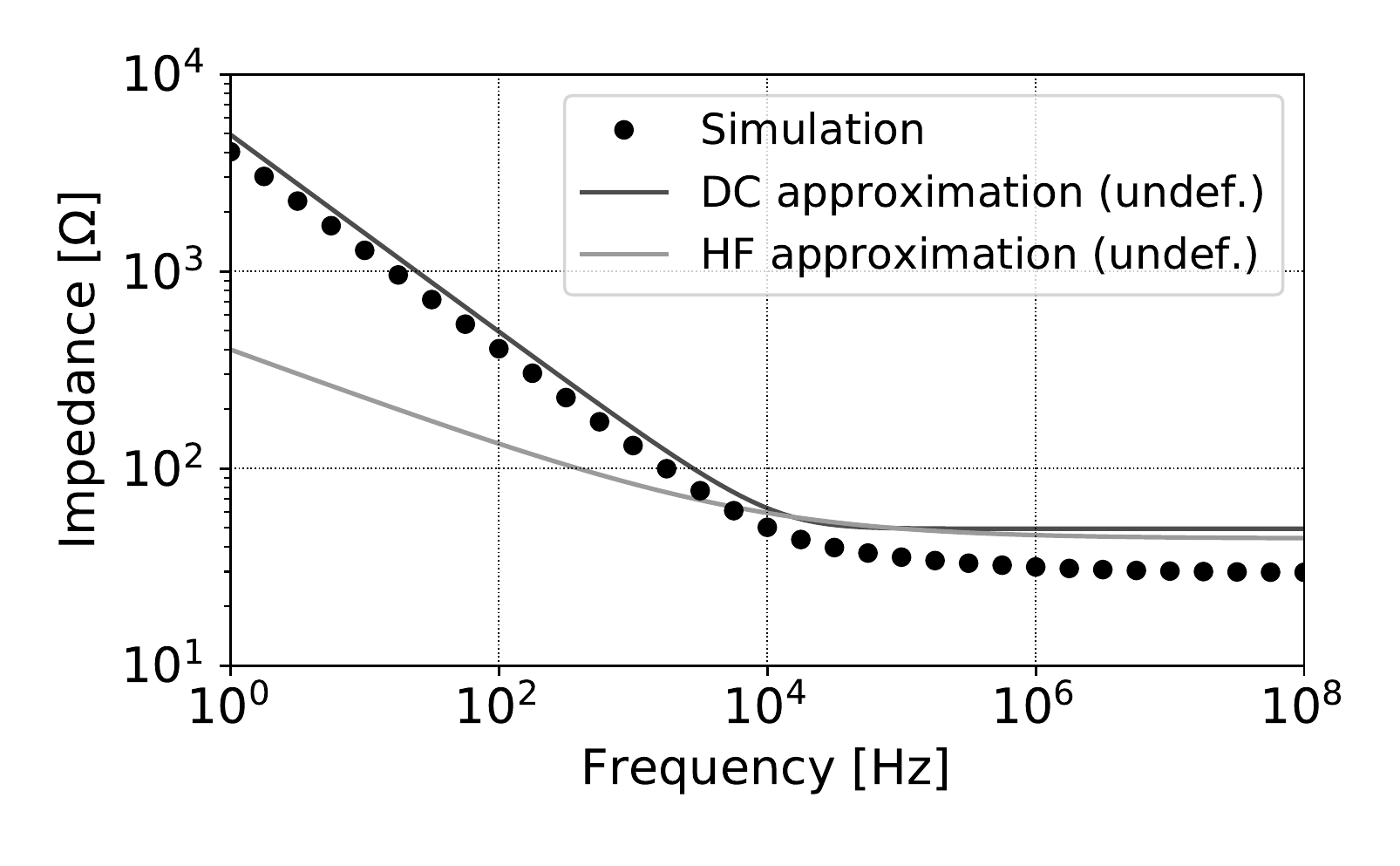}}
  \subfloat{\includegraphics[width=0.5\textwidth]{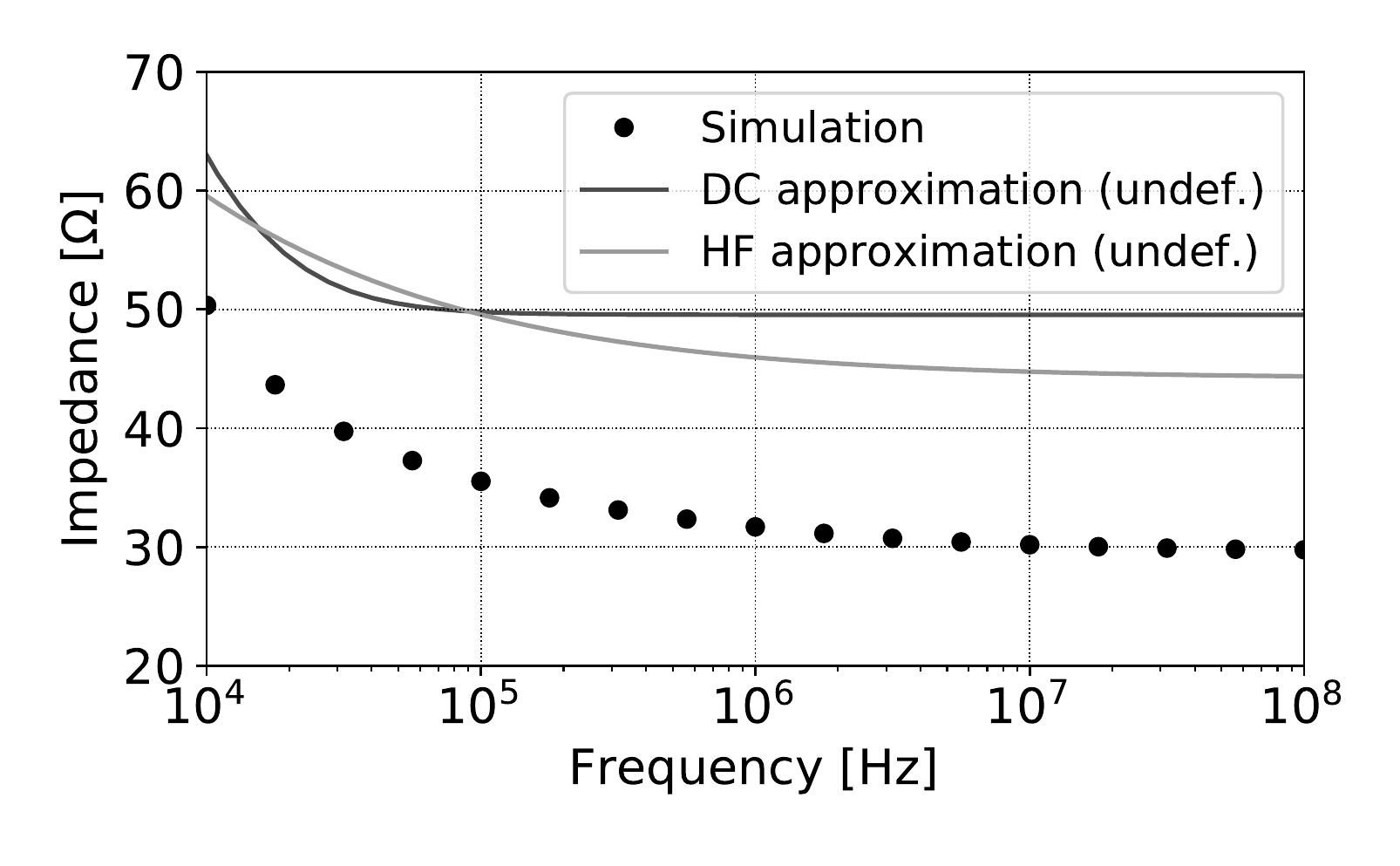}}
  \caption{\textbf{Transmission-line parameters of a deformed coaxial cable.} We show the transmission-line parameters and the characteristic impedance as defined by~\cref{eq:R,eq:L,eq:C,eq:G,eq:impedance} for a deformed coaxial cable. The analytic approximation describes the undeformed case. The number of grid points along one axis is $N = 512$ and the remaining parameters are $r_1 =\SI{0.48}{\milli\metre}$, $r_4 = \SI{0.8}{\milli\metre}$, $h = \SI{1.02}{\milli\metre}$, $t = \SI{0.15}{\milli\metre}$, $\epsilon_{\text{r},\text{dielectric}}^\prime = 2.25$, $\sigma_{\text{dielectric}}=0$, $\tan\left(\delta\right) = 10^{-3}$, $\epsilon_{\text{r},\text{copper}}^\prime = 1$~\cite{PhotonicCrystals}, $\sigma_{\text{copper}} = 5.98 \times 10^7 \frac{S}{m}$~\cite{johnson2003high}, $\epsilon_{\text{r},\text{air}}^\prime = 1.00059$~\cite{DielectricAir}, $\sigma_{\text{air}} = 0$.}
  \label{fig:coax deformed}
\end{figure*}

Next, we determine the transmission-line parameters of the undeformed coaxial cable according to~\cref{eq:R,eq:L,eq:C,eq:G}, and compare the obtained values with the analytical low and high frequency approximations. The analytical expressions for capacitance and conductance per unit length are~\cite{paul2008analysis}
\begin{align}
C &= \frac{2 \pi \epsilon_0 \epsilon^\prime_{\text{r}}}{\ln\left( \frac{r_2}{r_1}\right)}\,,\label{eq:C_analyt} \\
G &= \omega \tan\left(\delta\right) C \label{eq:G_analyt}\,.
\end{align}
We set $\epsilon^\prime_\text{r}=2.25$ and $\tan\left(\delta\right)=10^{-3}$. According to \cref{eq:C_analyt,eq:G_analyt}, the capacitance exhibits no frequency dependence and the conductance scales linearly with $\omega$. This behavior is captured by our simulations (see figure~\ref{fig:coax2}). Instead of setting $\tan\left(\delta\right)$ equal to a constant, it is also possible to use empirically determined frequency dependencies of $\tan\left(\delta\right)$.

The current densities in the conductors are frequency-dependent. For low frequencies, the current density is uniformly distributed within the conductor (DC approximation), and for high frequencies the current density is concentrated at the conductor surface (HF approximation). The skin effect is approximated by describing the conductors as hollow cylinders with their thickness being determined by the corresponding skin depth. In the low frequency regime, the resistance and conductance per unit length are~\cite{Hayt}
\begin{align}
R &= \frac{1}{\sigma \pi}\left( \frac{1}{r_1^2} + \frac{1}{r_3^2-r_2^2} \right)\,, \\
L &= \frac{\mu}{2 \pi} \left[ \ln\left( \frac{r_2}{r_1}\right) +  \frac{1}{4} \right. \nonumber \\
&\left.+\frac{1}{4 \left( r_3^2-r_2^2 \right)} \left(r_2^2-3 r_3^2+\frac{4 r_3^2}{r_3^2-r_2^2} \ln\left( \frac{r_3}{r_2} \right)\right) \right] \label{eq:L_analytic}
\end{align}
and for high frequencies
\begin{align}
R &= \sqrt{\frac{\mu \omega}{2 \sigma}} \frac{1}{2 \pi} \left( \frac{1}{r_1} + \frac{1}{r_2} \right)\,, \\
L &= \frac{\mu}{2 \pi} \ln\left( \frac{r_2}{r_1}\right) + \frac{\mu}{4 \pi} \frac{2}{\mu \omega \sigma} \left( \frac{1}{r_1} + \frac{1}{r_2} \right)\,.
\end{align}
We show the analytical approximations of $R$ and $L$ and the corresponding simulation results in figure~\ref{fig:coax2}. We find that the simulated transmission-line parameters are in good agreement with the analytic DC and HF approximations. In addition, our numerical results also describe the analytically inaccessible transition region between the DC and HF regime.

For large frequencies, deviations of the simulated resistance values from analytic theory occur. The reason is that very small grid spacings are necessary to resolve the current density and compute the resistance per unit length according to \cref{eq:jouleheating}. Another more computationally efficient possibility is to extrapolate the resistance according to the scaling of $R\left(\omega\right)\sim \sqrt{\omega}$ for $\omega \rightarrow \infty$. If one is only interested in the impedance, it is also possible to neglect resistance-related deviations since
\begin{equation}
Z_0\left(\omega\right) = \sqrt{\frac{L}{C}}\quad \text{for} \quad \omega \rightarrow \infty\,.
\label{eq:HF_Z0}
\end{equation}
\subsection{Deformed coaxial transmission line}
To study the influence of deformations on transmission-line parameters, we now compress the coaxial transmission line between two rigid parallel plates (see figure~\ref{fig:cross section coax}). The circumference is invariant under the applied deformation, so
\begin{equation}
2 \pi r_2 = 2 \pi r_4 + 4 h\,,
\end{equation}
where $r_2$ is the inner radius of the outer conductor of the undeformed cable, $r_4$ is the radius of the upper and lower half circles of the deformed cable, and $h$ is the distance of the centers of the half circles from the center of the inner conductor (see figure~\ref{fig:cross section deformed coax}).

We consider a deformation with $r_4=\SI{0.8}{\milli\metre}$ and $h=\SI{1.02}{\milli\metre}$, and compare the transmission-line parameters of this deformed coaxial cross-section with those of the undeformed reference. We show the corresponding results in figure~\ref{fig:coax deformed}. We find that resistance and conductance are almost unaffected by the deformation. However, capacitance and inductance differ significantly when compared to the undeformed reference. In the high frequency regime, the impedance of the undeformed coaxial cable is about 50 \% larger than the impedance of the deformed one. According to \cref{eq:reflection_coeff}, this corresponds to a reflection coefficient of $\Gamma=-0.2$ at the interface between the undeformed and deformed segments. We outline in section~\ref{sec:experiment}, that such reflections are detectable with a time-domain reflectometer (TDR). In addition to variations in the cross section, it is also possible to incorporate spatial permittivity variations that result from applied mechanical strains.
\subsection{Comparison with experimental data}
\label{sec:experiment}
\begin{figure*}[h!]
  \centering
  \subfloat[Deformed transmission line (simulation).]{\label{fig:L45 simulation}\includegraphics[angle=0,clip,width=0.46\textwidth, trim={0.2cm 5.4cm 0.2cm 5.1cm}]{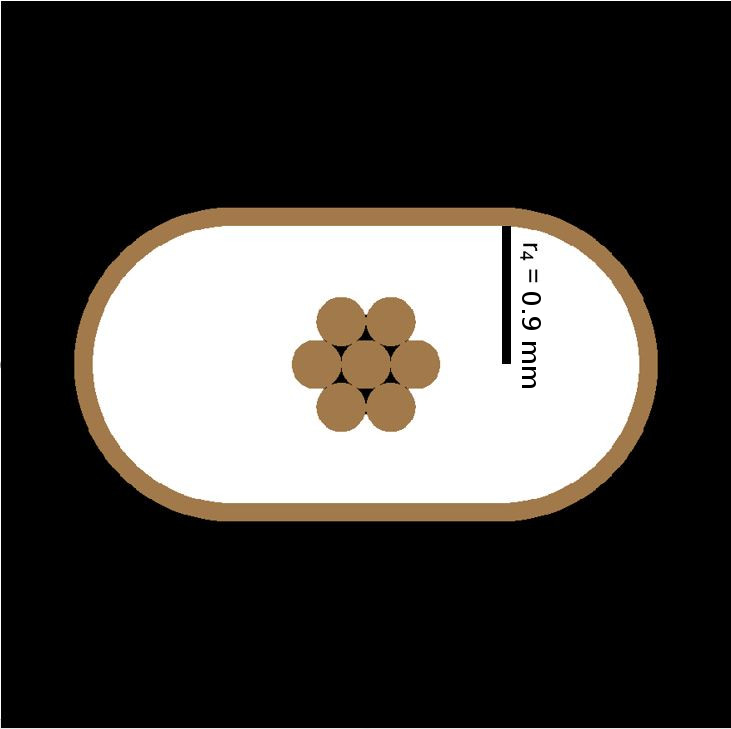}}
  \hspace*{\fill}
  \subfloat[Deformed transmission line (microscope).]{\label{fig:L45 microscope}\includegraphics[angle=0,clip,width=0.46\textwidth]{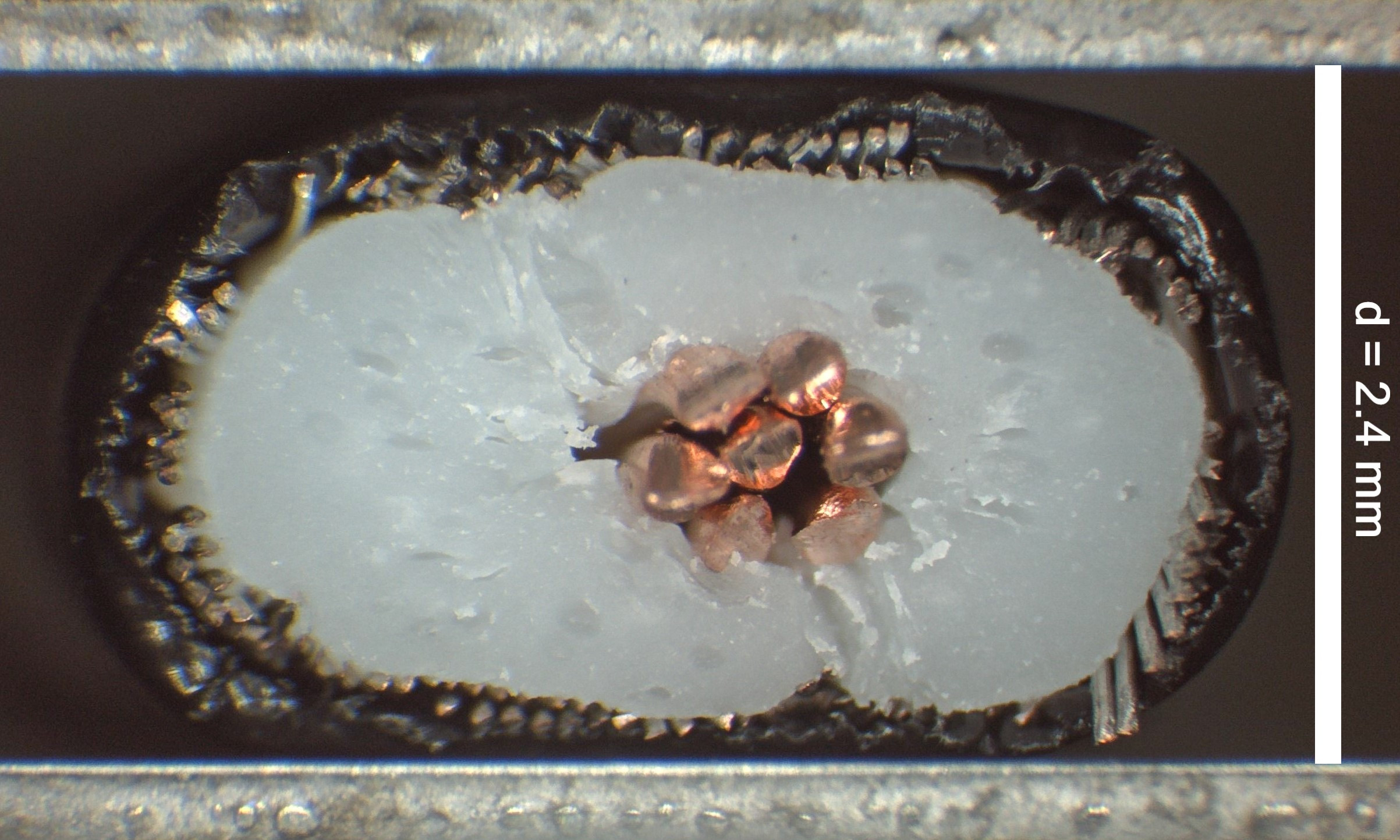}}
  \hfill
  \subfloat[TDR measurement and simulation.]{\label{fig:L45 TDR}\includegraphics[width=0.5\textwidth]{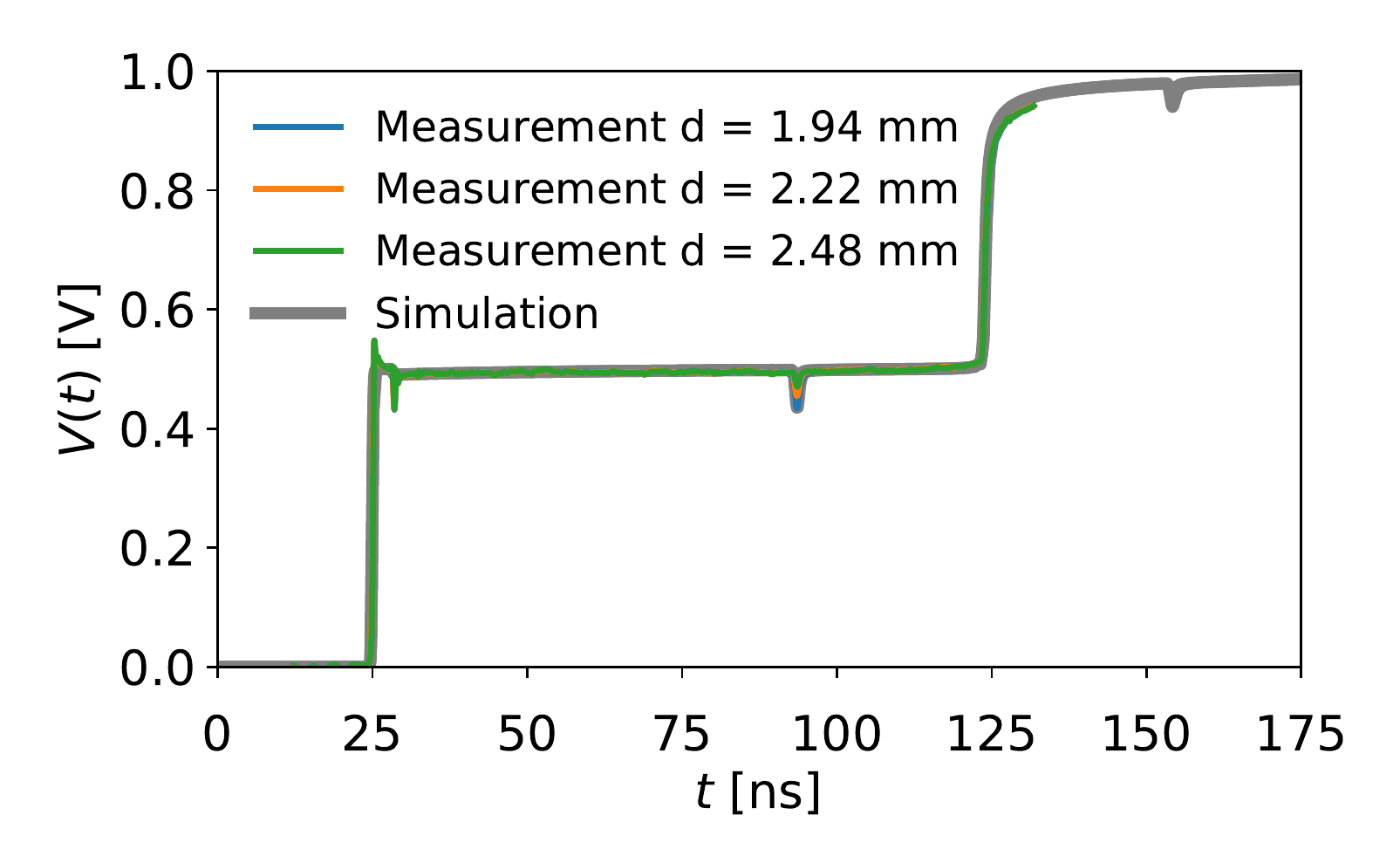}}
  \subfloat[TDR measurement and simulation (zoomed).]{\label{fig:L45 TDR zoom}\includegraphics[width=0.5\textwidth]{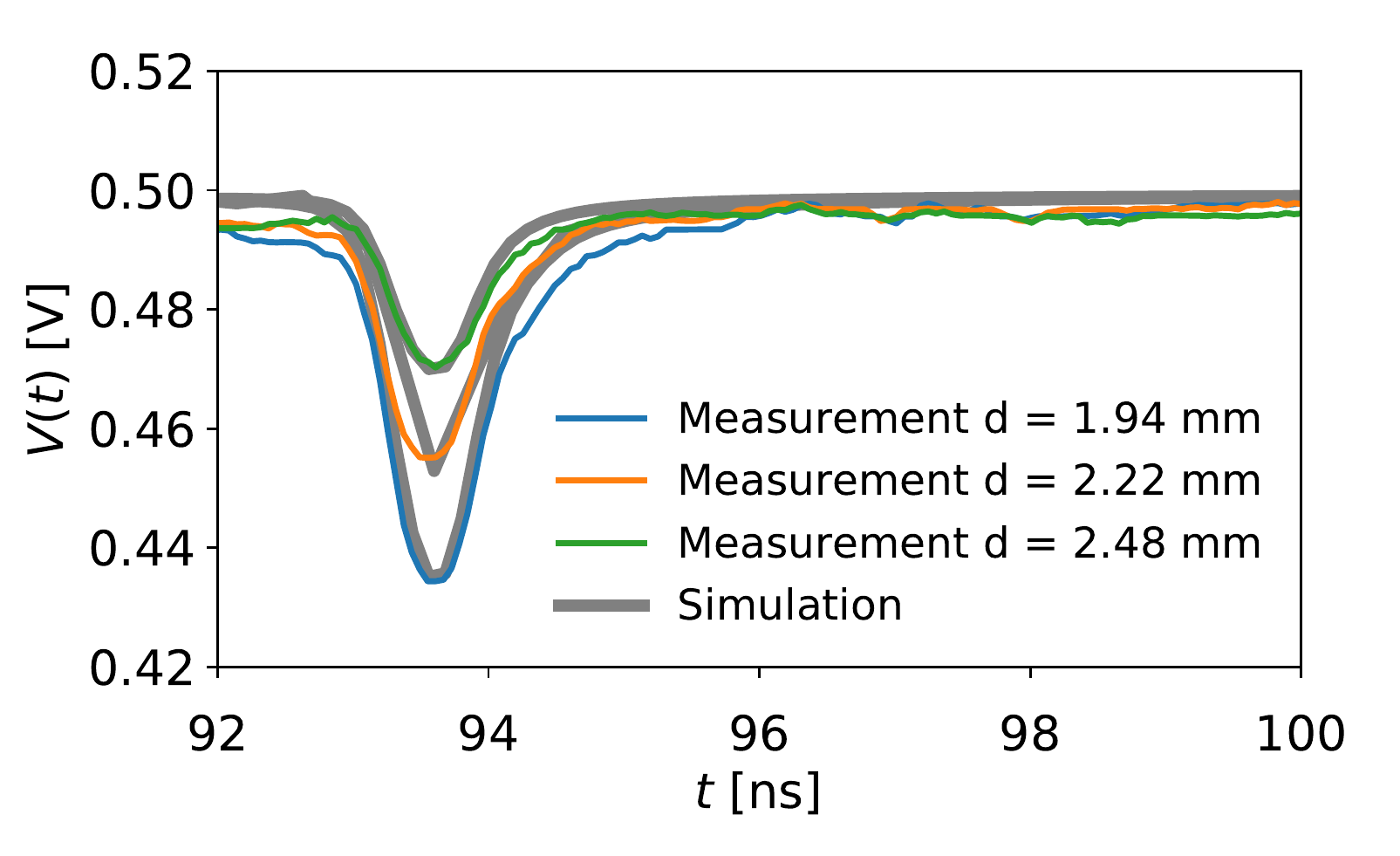}}
  \caption{\textbf{Comparison of simulations with experimental data.} The upper panels show the deformed coaxial transmission-line cross sections (simulation and experiment). The inner conductor consists of seven copper wires of radius $r_1 = \SI{0.16}{\milli\metre}$, the shield is a braiding of tinned copper wires with a thickness of $t = \SI{0.15}{\milli\metre} $, and the dielectric material is a thermoplastic elastomer with a relative permittivity of $\epsilon_{\text{r},\text{dielectric}}^{\prime} = 2.14$ and $\epsilon_{\text{r},\text{dielectric}}^{\prime\prime} = 4.28 \times 10^{-3}$ and a radius of $r_2 = \SI{1.45}{\milli\metre}$. The cable jacket has a thickness of $\SI{0.40}{\milli\metre}$. The simulation neglects that the shield is made of multiple tinned copper wires but approximates the cable by a non-braided copper shield of thickness $t = \SI{0.12}{\milli\meter}$. In the lower panels, we show the corresponding TDR measurements for the described cable of length $\SI{9.70(1)}{\metre}$. We apply the shown deformation $\SI{6.98(1)}{\metre}$ away from the TDR input with a vise of length $\SI{6}{\centi\metre}$. We denote the distance between the two parallel plates of the vise by $d$.}
  \label{fig:L45 deformed}
\end{figure*}
Our simulations show that the impedance of a deformed coaxial transmission line is significantly smaller as compared to the undeformed case (see figure~\ref{fig:coax deformed}). To test our numerically obtained transmission-line parameters and impedance values, we experimentally analyze the effect of deformations on the TDR profile of a deformation-sensitive coaxial cable. TDR measurements are based on the injection of a test signal (\textit{e.g.}, a step function) into a transmission line. Reflections of the injected signal occur at impedance discontinuities. Each reflection in the transmission line leads to a wave which travels back to the TDR where the resulting voltage differences and corresponding time stamps are monitored. These monitored quantities contain information about the spatial distribution of reflection coefficients and impedances along the transmission line.
\begin{table}[h!]
\centering
\begin{tabular}{c|c|c}
$d \left[\SI{}{\milli\metre}\right]$ & $r_4 \left[\SI{}{\milli\metre}\right]$ & $h \left[\SI{}{\milli\metre}\right]$ \\ \hline
2.48(3) & 0.95(1) & 0.79(1) \\
2.22(3) & 0.85(1) & 0.93(1) \\
1.94(3) & 0.75(1) & 1.10(1) \\
\end{tabular}
\caption{For each measurement and simulation in figures~\ref{fig:L45 TDR} and \ref{fig:L45 TDR zoom}, we summarize the corresponding cable geometry parameters. Parameters $r_4$ and $h$ determine the simulated deformation according to figure~\ref{fig:cross section deformed coax} and $d$ is the distance between the two parallel plates of the used vise.} 
\label{tab:geometry}
\end{table}
For our experiments, we use a Campbell Scientific TDR100. The considered coaxial transmission line has a length of $\SI{9.70(1)}{\metre}$ and its dielectric material is a thermoplastic elastomer with a relative permittivity of $\epsilon_{\text{r},\text{dielectric}}^{\prime} = 2.14$. The number in parentheses is the uncertainty in the last digit. With a vise of length $\SI{6.0(1)}{\centi\metre}$, we apply a deformation $\SI{6.98(1)}{\metre}$ away from the TDR. We denote the distance between the two parallel plates of the vise by $d$. We show an illustration of the deformed transmission lines that we considered in our simulations and experiments in figures~\ref{fig:L45 simulation} and \ref{fig:L45 microscope}. For the considered transmission line, we did not observe any traces of plastic-deformation effects in the TDR curves. However, depending on the used dielectric and cable jacket materials, plastic deformations may occur and affect measurements. Under the quasi-TEM assumption, the transmission line is composed of undeformed and deformed segments. We have to compute the frequency dependence of the transmission-line parameters for both cross sections only once to simulate the corresponding TDR profile according to the telegraphers equations as defined in \cref{eq:telegraphers1,eq:telegraphers2}. To solve \cref{eq:telegraphers1,eq:telegraphers2}, we employ a method as described in \cite{d2017transmission}. We again note that we only consider geometric effects in our simulations, and that we do not account for strain-induced spatial variations of the permittivity within the dielectric material. To test the validity of our simulation approach, we compare the simulated TDR profiles with experimentally obtained ones. We summarize the parameters that we use in our simulations in table~\ref{tab:geometry}. In figures~\ref{fig:L45 TDR} and \ref{fig:L45 TDR zoom}, we show the comparison between our simulations and corresponding experiments. These results clearly show that our framework is able to reproduce the observed TDR profiles. 
Determining the dependence of the voltage minimum on the plate distance $d$ in figure~\ref{fig:L45 TDR zoom} requires to numerically solve \cref{eq:telegraphers1,eq:telegraphers2} for a given set of transmission-line parameters/impedances.
A closed-form analytical expression can only be obtained for the impedance of concentrically-deformed transmission lines (\textit{i.e.}, coaxial lines of different diameter)~\cite{LiThAb}. In the case of concentric deformations, we can obtain the characteristic impedance from \cref{eq:C_analyt,eq:G_analyt,eq:L_analytic,eq:HF_Z0}:
\begin{equation}
Z_0(\omega)=\frac{1}{2 \pi} \sqrt{\frac{\mu_0 \mu_r}{\epsilon_0 \epsilon_r}} \ln\left(\frac{r_2}{r_1}\right)\quad \text{for} \quad \omega \rightarrow \infty\,.
\end{equation}
To describe the effect of general deformations on $Z$, it is necessary to use a numerical framework such as the one we propose in this work.
In particular, we find that the simulations are able to capture the functional dependence of voltage minimum and plate distance $d$ in the deformed regions. The experimentally observed TDR profiles are therefore consistent with our numerical framework for the simulation of arbitrarily-shaped transmission-line cross sections.
\section{Conclusion}
\label{sec:Discussion}
We have introduced a numerical framework to simulate arbitrarily-shaped transmission lines over a wide frequency range. Our method is based on solving a sparse system of linear equations and thus is suitable for fast computations. The numerically computed transmission-line parameters and electromagnetic field values agree well with corresponding analytical predictions. We considered the effect of geometric deformations, but our framework can be extended to also account for strain-induced permittivity variations. We tested the proposed framework by comparing simulation results with corresponding experimental data and found good agreement between the model predictions and the experimentally observed behavior.

Our study may improve continuous sensing applications for the measurement of deformations and mechanical strains based on strain-sensitive transmission lines. Instead of mapping observed impedance values to concentric deformations and thus neglecting any further structural information as suggested in~\cite{LiThAb}, our method also captures changes in the cross-section geometry. Furthermore, our framework is also useful to determine analytically inaccessible transmission-line parameters and impedances for complex arrangements of conductor, insulator, and shielding materials. 

A possible extension of our work is to combine our framework with an inversion algorithm and determine estimates of the most likely deformations and mechanical strains given a certain TDR observation. This can contribute to improvements in corrosion detection~\cite{liu2002corrosion} and monitoring fracture propagation in concrete structures~\cite{LiThAb,chen2003continuous}.
When embedding transmission-line sensors in concrete structures, it is important to examine strain transfer characteristics at the interface between cable jacket and surrounding material. Similar to fibre-optic sensors~\cite{santos2014handbook}, the mechanical and physicochemical properties of the materials at the interface between sensor line and surrounding material (\textit{e.g.}, cable jacket material) may substantially influence the sensor performance. Recent advances in strain-transfer theory~\cite{wang2018strain} may provide a possibility to correctly interpret measurement data, even in the presence of strain-transfer loss at the interface.

To enhance the long-term sensing performance, it is important to limit material degradation by using materials with physicochemical properties that are suitable for a given application~\cite{santos2014handbook}. For instance, photochemical and oxidative degradation (\textit{i.e.}, chemical aging) can be limited by lowering the influence of air and light on the sensor. In addition, one should also consider the effect of the change of physical properties over time (\textit{i.e.}, physical aging) that may occur in the used polymer blends~\cite{hodge1995physical}.
If transmission-line sensors are tailored to a specific application, previous studies suggest that they can stay in service for decades~\cite{LiThAb}.
Depending on the used dielectric material, it may also be important to account for the influence of temperature variations on the measurement results if the permittivity temperature coefficient is large enough. The dielectric material of the transmission line we consider has a temperature coefficient of about $10^{-4}~\text{K}^{-1}$~\cite{harrop1969temperature}, so the effect of temperature variations on the impedance is negligible. In addition to permittivity changes, temperature variations may also affect the coating material and strain transfer characteristics~\cite{wang2019strain}.
\section*{Acknowledgements}
We thank Dani Or and Daniel Breitenstein for the possibility to use their Campbell Scientific TDR100 time-domain reflectometer and Joshua LeClair for helpful comments and discussions. We also thank the LEONI AG for providing a sample of a pressure-sensitive coaxial transmission line.
%
%
%
\section*{References}
\providecommand{\newblock}{}

\end{document}